\documentclass[useAMS,usenatbib,usegraphicx]{mn2e}

\usepackage{times}

\def\aj{AJ}             % Astronomical Journal
\def\araa{ARA\&A}       % Annual Review of Astron and Astrophys
\def\apj{ApJ}           % Astrophysical Journal
\def\apjl{ApJ}          % Astrophysical Journal, Letters
\def\apjs{ApJS}         % Astrophysical Journal, Supplement
\def\aap{A\&A}          % Astronomy and Astrophysics
\def\mnras{MNRAS}       % Monthly Notices of the RAS
\def\pasp{PASP}         % Publications of the ASP
\def\skytel{S\&T}       % Sky and Telescope
\def\nat{Nature}        % Nature
\def\lesssim{\mathrel{\hbox{\rlap{\hbox{\lower4pt\hbox{$\sim$}}}\hbox{$<$}}}}
\def\gtrsim{\mathrel{\hbox{\rlap{\hbox{\lower4pt\hbox{$\sim$}}}\hbox{$>$}}}}

\newcommand{\reduceme}{\mbox{R\raisebox{-0.35ex}{E}D%
\hspace{-0.05em}\raisebox{0.85ex}{uc}\hspace{-0.90em}%
\raisebox{-.35ex}{{m}}\hspace{0.05em}E}}

\def\oiii{[O\,{\sc iii}]}
\def\halpha{\hbox{H$\alpha$}}
\def\hbeta{\hbox{H$\beta$}}
\def\ni{\hbox{[N\,{\sc i}]}}
\def\nii{\hbox{[N\,{\sc ii}]}}

\def\kms{km~s$^{-1}$}
\def\Msun{${\rm M_{\odot}}$}

\def\sauron{{\tt SAURON}}
\def\oasis{{\tt OASIS}}
 
\begin{document}
 
\author[J.\ Falc\'on-Barroso et al.]
{Jes\'us Falc\'on-Barroso$^{1,2}$\thanks{E-mail:jfalcon@strw.leidenuniv.nl}, 
Reynier F.\ Peletier$^{1,3}$, 
Eric Emsellem$^{3}$,
Harald Kuntschner$^{4}$,
\newauthor
Kambiz Fathi$^{1,3}$, 
Martin Bureau$^{5}$\thanks{Hubble fellow},
Roland Bacon$^{3}$,
Michele Cappellari$^{2}$,
\newauthor
Yannick Copin$^{6}$,
Roger L.\ Davies$^{7}$,
Tim de Zeeuw$^{2}$
\\$^1$ School of Physics \& Astronomy. University of Nottingham. Nottingham, NG7~2RD, United Kingdom
\\$^2$ Sterrewacht Leiden, Niels Bohrweg~2, 2333~CA, Leiden, The Netherlands
\\$^3$ CRAL, Observatoire de Lyon, F-69561 St-Genis Laval cedex, France
\\$^4$ Space Telescope European Coordinating Facility, European Southern Observatory, Karl-Schwarzschild-Str.~2, 85748 Garching, Germany
\\$^5$ Columbia Astrophysics Laboratory, 550 West 120$^{th}$ Street, 1027 Pupin Hall, MC~5247, New York, NY~10027
\\$^6$ Institut de physique nucleaire de Lyon, 4 rue Enrico Fermi, 69622 Villeurbanne Cedex, France
\\$^7$ University of Oxford, Astrophysics, Keble Road, Oxford, OX1~3RH, United Kingdom\\}
\date{Accepted ... Received ...; in original form}

\title[Formation and evolution of S0 galaxies]{Formation and evolution of S0 galaxies: a \sauron\ case study of NGC\,7332}
\maketitle

%%%%%%%%%%%%%%%%%%%%%%%%%%%%%%%%%%%%%%%%%%%%%%%%%%%%%%%%%%%%%%%%%%%%%%%%%%%%%%%%%%%%%%%%%%%%%%%%%%%%%
\begin{abstract}
  We present \sauron\ integral-field observations of the S0 galaxy  NGC\,7332.
  Existing broad-band ground-based and HST photometry reveals a double disk
  structure and a boxy bulge interpreted as a bar  viewed close to edge-on. The
  \sauron\ two-dimensional stellar kinematic maps  confirm the existence of the
  bar and inner disk but also uncover the  presence of a cold counter-rotating
  stellar component within the central  $250$~pc. The \hbeta\ and \oiii\ emission
  line maps show that the  ionised gas has a complex morphology and kinematics,
  including  both a component counter-rotating with respect to the stars and a
  fainter co-rotating one. Analysis of the absorption line-strength maps show that
  NGC\,7332 is young everywhere. The presence of a large-scale bar can explain
  most of those properties, but the fact that we see a  significant amount of
  unsettled gas, together with a few peculiar  features in the maps, suggest that
  NGC\,7332 is still evolving. Interactions  as well as bar-driven processes must
  thus have played an important  role in the formation and evolution of NGC\,7332,
  and presumably of S0  galaxies in general.
\end{abstract}
 

%%%%%%%%%%%%%%%%%%%%%%%%%%%%%%%%%%%%%%%%%%%%%%%%%%%%%%%%%%%%%%%%%%%%%%%%%%%%%%%%%%%%%%%%%%%%%%%%%%%%%
\section{Introduction}
\label{Sec:Introduction}

The nature of S0 galaxies has been a matter of controversy since their  appearance
in the first morphological classification schemes 
\citep{hubble36,devac59,vdbergh60a,vdbergh60b} and many authors refer to  them as
normal spirals stripped of their gas  \citep[e.g.,][]{sandage78,dressler80}.
Studies of S0 galaxies in clusters  suggest that interactions between galaxies can
lead to the loss of  interstellar matter in the disk and convert spirals into S0s.
Other  theories favour gas loss through ram-pressure stripping from the 
intergalactic medium \citep{gunn72} or via galactic winds \citep{faber76b},  or
assert that S0 galaxies are primordial galaxies which have entirely  consumed
their gas due to a high star formation rate \citep{larson80}.

When studying the formation of S0 galaxies, one is often drawn to examine the
order of events during galaxy assembly. Early formation models favour the
formation of the bulge before the disk, which is supported by several photometric
studies \citep[e.g.,][]{caldwell83,terndrup94,pb96,peletier99} and by the fact
that S0s behave like ellipticals in the widely known scaling relations (i.e.\
color-magnitude, Mg$_2$-$\sigma$ or fundamental plane).  Numerical simulations
raised the possibility of an alternative scenario \citep[e.g.,][]{pfenniger93}, in
which bulges are formed via secular evolution processes after the disk 
\citep[e.g.,][]{sw93,courteau96,carollo99,bureau99b,merrifield99}.  Although such
studies naturally focus on the two major components, i.e.\ the large-scale disk
and the bulge, galaxies often show more structures,  e.g.\ central disks, stellar
kinematically decoupled components (KDCs)  and multiple gaseous components. These
are commonly observed in galaxies  all along the Hubble sequence and are often
used to emphasize the role of  mergers during the formation of early-type
galaxies  \citep[][]{toomre77,wr78,schweizer86,fi88}. 

In the last $15$ years, evidence for the presence of KDCs in early-type  galaxies
(e.g.,\ NGC\,4406, \citealt{bender88}; NGC\,4365, \citealt{wbm88})  has usually
been revealed through long-slit measurements. This can introduce  a priori
assumptions about the geometry of the central structures, and  several works have
tried to overcome this limitation by mapping the galaxies  with multiple slits at
different positions angles  \citep{fisher94,fisher97,ss99,michele02}. However, the
arrival of a new  generation of integral-field units (IFUs) providing contiguous
field-of-views  (FOVs), such as \oasis\ (e.g., NGC\,4621, \citealt{wernli02}) or
\sauron\  \citep{bacon01,tim02}, offers a much more efficient way to detect KDCs. 

Early \sauron\ results \citep{bacon01,tim02} reveal a variety of  structures much
richer than usually recognized in early-type galaxies.  It is therefore important
to use the information contained in these  features to elucidate the key processes
at work during galaxy  formation. The galaxies concerned can become the ideal
benchmarks  against which to test the predictions of galaxy formation and
evolution  theories. NGC\,7332, the galaxy studied in this paper, may be such a
keystone. 

NGC\,7332, a boxy edge-on S0 galaxy\footnote{The galaxy is classified  S0p in the
RC3 \citep{RC3} due to its boxy bulge \citep{sandage61}.},  was studied
extensively in the past. The galaxy is mainly known for a  bright counter-rotating
and a faint co-rotating \oiii\ gas component with  respect to the stars
\citep{bertola92,fisher94}. Those gas structures were  confirmed by
\citet{plana96}, who mapped the \halpha\ emission via  Fabry-Perot observations.
NGC\,7332 exhibits a rather regular broad-band  morphology \citep{pb97} and its
colours are somewhat bluer than those of  elliptical galaxies of the same
luminosity. Spectral analysis of the central  regions reveals a
luminosity-weighted age of about $6$~Gyr \citep{vazari99}. 

In this paper, we use \sauron\ observations to characterize new and known features
and improve our understanding of the morphology and dynamics of NGC\,7332. We
focus the analysis on the distribution and kinematics of  the stellar and gaseous
components, but we also discuss the corresponding  \sauron\ line strengths maps.
We present strong evidence for the presence  of a central KDC and complex
two-dimensional (2D) gas kinematics. 

The main characteristics of NGC\,7332 are provided in Table~\ref{Tab:Params}.
Section~\ref{Sec:Obsdata} summarises the observations discussed in this paper and
their reduction. A photometric analysis and comparison with the literature is
presented in Section~\ref{Sec:Phot}. In Sections~\ref{Sec:Skinem} and
\ref{Sec:Gkinem}, we present and discuss the kinematic maps of stars and gas, and
describe the methods used to disentangle their respective spectral contributions.
The line-strength analysis is presented in Section~\ref{Sec:Popu}. We finally
discuss possible formation and evolution scenarios for NGC\,7332 in
Section~\ref{Sec:Discussion} and summarise our conclusions in 
Section~\ref{Sec:Conclusions}.

%%%%%%%%%%%%%%%%%%%%%%%%%%%%%%%%%%%%%%%%%%%%%%%%%%%%%%%%%%%%%%%%%%%%%%%%%%%%%%%%%%%%%%%%%%%%%%%%%%%%%
\section{Observations \& Data Reduction}
\label{Sec:Obsdata}

%----------------------------------------------------------------------------------------------------
\subsection{\sauron\ observations \& data reduction}
\label{Sec:SauronObs}

We observed NGC\,7332 with the integral field spectrograph \sauron\ attached  to
the $4.2$-m William Herschel Telescope (WHT) of the Observatorio del Roque  de los
Muchachos at La Palma, Spain, on 13 Oct 1999. We obtained $4$ largely  overlapping
exposures of $1800$\,s each, producing more than $1500$ spectra per exposure,
including $146$ sky spectra $1\farcm9$ away from the main field. \sauron\ delivers
a spectral resolution of $4.3$~\AA\ (FWHM) and covers the narrow spectral range
$4810-5350$~\AA\  ($1.1$~\AA~pixel$^{-1}$). The spatial sampling of individual
exposures is performed by an array of square lenses of $0\farcs94$, providing a
FOV of $\approx33\arcsec\times41\arcsec$. The seeing at the time of the
observations was stable at about $1\farcs1$ (FWHM). Flux, velocity and
line-strength standard stars were observed during the same observing run for
calibration purposes. Arc lamp exposures were taken before and after each target
frame for wavelength calibration. A tungsten lamp exposure was also taken at the
end of the night in order to build the mask allowing us to extract the data from
the CCD frames.

We followed the procedures described in \citet{bacon01} for the extraction,
reduction, and calibration of the data, using the specifically designed {\tt
XSauron} software. The sky level was measured with the help  of the dedicated sky
lenses and subtracted from the target spectra. We merged the $4$ individual
extracted  datacubes by spatially resampling the spectra onto a common squared
grid.  The dithering of individual exposures (with an original spatial sampling
of  $0\farcs94\times0\farcs94$) enabled us to sample the merged datacube onto 
$0\farcs8\times0\farcs8$ pixels. The resulting merged mosaic FOV is 
$\approx34\arcsec\times46\arcsec$ with a total of $1881$ spectra. We then  binned
our final datacube using the Voronoi 2D binning algorithm  of \citet{michele03} to
create compact bins with a minimum signal-to-noise  ratio $S/N$ of $60$ per pixel.
Most spectra have $S/N$ in excess  of $60$, however, so that most of the original
spatial elements remain  un-binned (e.g., $[S/N]_{\rm max}\approx300$ in the
central lens).

\begin{table}
\begin{center}
\caption{Properties of NGC\,7332}
\label{Tab:Params}
\begin{tabular}{lrl}
\hline
Parameter                       & Value                & Source \\
\hline
Morphological Type              & S0~pec               & \citealt{RC3} \\
M$_R$ [mag]                     & -21.86               & \citealt{bp94} \\
$B-R$ [mag]                     & 1.40                 & \citealt{bp94} \\
Outer Ellipticity               & 0.75                 & \citealt{apb95} \\
Distance Modulus [mag]          & 31.81                & \citealt{tonry01} \\
Distance scale [pc / arcsec]    & 111.6                &  \\
\hline
\end{tabular}
\end{center}
\end{table}

%----------------------------------------------------------------------------------------------------
\subsection{STIS spectroscopy \& data reduction}
\label{Sec:STISObs}
We also retrieved Space Telescope Imaging Spectrograph (STIS) data from the Hubble
Space Telescope (HST) archive at the Space  Telescope Science
Institute\footnote{Based  on observations made with the NASA/ESA Hubble  Space
Telescope, obtained from the data archive  at the Space Telescope Science
Institute. STScI  is operated by the Association of Universities  for Research in
Astronomy, Inc., under NASA  contract NAS 5-26555.} (STScI) to obtain high spatial
resolution spectroscopy of NGC\,7332's centre. The data form part of the
observations of proposal ID~7566 by R.\ Green, covering the Ca\,{\sc ii} region
between $8282$ and $8835$~\AA. The G750M grating was used together with the
$52\arcsec\times0\farcs2$ slit aperture, providing a dispersion of 
$0.56$~\AA~pix$^{-1}$ and a spatial sampling of $0\farcs05$~pix$^{-1}$.  We
retrieved the data reduced by the pipeline, making use of the best  calibration
files available at the time, but corrections for the  effect of fringing and
cosmic ray removal were still required. Fringing  corrections were applied using
the prescriptions in \citet{goud97},  whereas we used the \reduceme\ package
\citep{cardiel99} to perform  the cosmic ray rejection. A K0III star (HR7615) from
the same data  set was also retrieved and underwent the same reduction processes. 
It was used as a template for kinematical measurements.

\begin{figure}
\centering
{\includegraphics[angle=00, width=1.\linewidth]{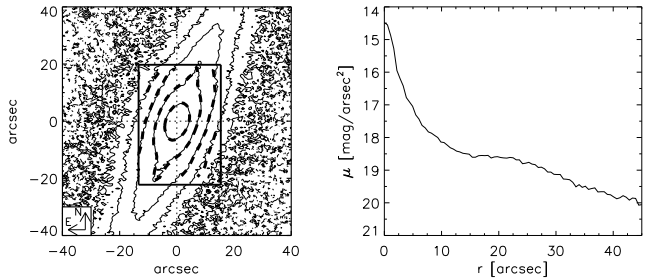}}
\caption{\sauron\ FOV (thick box) superposed on the
  $I$-band image of \citet{pb97} (left panel). The orientation is 
  indicated in the bottom-left corner and is the same as for the 
  \sauron\ maps shown in Figures~\ref{Fig:Skin}-\ref{Fig:Indices}. 
  Contour levels are separated by $1$ mag. The $I$-band major-axis 
  surface brightness profile from \citet{pb97} is shown in the right 
  panel.}
\label{Fig:FoV}
\end{figure}

%----------------------------------------------------------------------------------------------------
\subsection{HST \& ground-based imaging}
\label{Sec:Imag}

Imaging data were retrieved from several archives to perform a structural
analysis. We retrieved a set of Wide-Field Planetary Camera (WFPC-1) images from
the HST archive at STScI from the observing program of S.\ Faber (Prop.\ ID~2600,
see \citealt{lauer95}). The  observations were made using the F555W filter
($V$-band). The data  set consists of $3$ target frames with exposure times of
$35$, $140$  and $140$~s, respectively, and spatial sampling
$0\farcs044$~pixel$^{-1}$.  We combined the individual exposures and rejected
cosmic rays with the  routine {\it crrej}, available within the IRAF\footnote{IRAF
is distributed  by the National Optical Astronomy Observatories, which are
operated  by the Association of Universities for Research in Astronomy, Inc., 
under cooperative agreement with the National Science Foundation.}  package, thus
resulting in a cosmic ray-cleaned merged image. We also made  use of a
fully-reduced ground-based Cousins $I$-band image from \citet{pb97},  taken with
the Isaac Newton Telescope (INT) located in the Observatorio del  Roque de los
Muchachos in La Palma, Spain. The exposure time was $200$~sec  and the seeing
$1\farcs0$. We overlay in Figure~\ref{Fig:FoV} the contours  from this $I$-band
image and the FOV and contours of the \sauron\ reconstructed  total intensity map
(obtained by summing the data in wavelength). Finally,  the photometric analysis
performed by \citet{fisher94} from an $R$-band image,  taken in $1\farcs3$ seeing,
was also used.

%%%%%%%%%%%%%%%%%%%%%%%%%%%%%%%%%%%%%%%%%%%%%%%%%%%%%%%%%%%%%%%%%%%%%%%%%%%%%%%%%%%%%%%%%%%%%%%%%%%%%
\section{Photometric Analysis}
\label{Sec:Phot}

\begin{figure}
\centering
\includegraphics[angle=00,width=1.\linewidth]{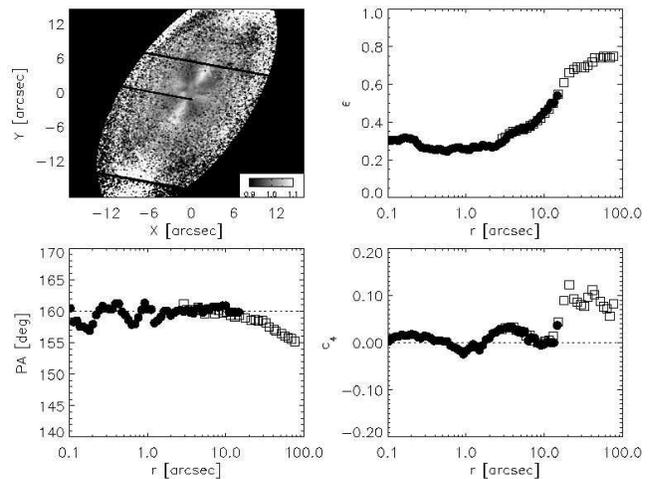}
\caption{Photometric analysis of NGC\,7332. The ratio of the WFPC-1 image
and a model image from our best ellipse fit is shown in the top left
panel. Position angle (PA), ellipticity ({\large $\epsilon$}) and boxiness ($c_4$) 
profiles along the major-axis radius ($r$) are shown in the other panels. 
Solid circles represent our WFPC-1 analysis whereas open squares 
show the profiles obtained by \citet{fisher94}. The formal errors on the different 
parameters (on the WFPC-1 data) are 0.60, 0.005 and 0.004 for 
the PA, {\large $\epsilon$} and $c_4$ respectively.}
\label{Fig:centdisk}
\end{figure}

We analysed the WFPC-1 data in order to study morphological sub-structures  in
NGC\,7332. A comparison of our analysis with that of \citet{fisher94}  from an
$R$-band image is shown in Fig.~\ref{Fig:centdisk} and the agreement  in the
overlapping region is excellent. This comparison is possible because  the color
gradients in this galaxy are small \citep{pb97} and the emission  lines can be
neglected (e.g., \halpha\ and \nii\ in the $R$-band image).

The isophotal analysis of the HST image reveals a number of maxima and minima in
the $c_4$ parameter, which describes deviations of the isophotes  from pure
ellipses (positive $c_4$ indicates disky isophotes and negative  $c_4$ boxy
isophotes; see e.g., \citealt{carter78}). Positive values are  present in the
inner $0\farcs5$ but may be strongly influenced by the  halo of the pre-COSTAR
WFPC-1 point-spread-function (PSF). Positive $c_4$  values are also observed in
the region $1\farcs5\lesssim \rm{r} \lesssim7\arcsec$,  where the maximum reaches
about 3 per cent. The isophotes are then elliptical  or slightly boxy at
intermediate radii ($7\arcsec\lesssim \rm{r} \lesssim12\arcsec$)  and again
strongly disky at large radii ($\rm{r} \gtrsim12\arcsec$). A complementary 
representation is shown in Figure~\ref{Fig:centdisk} (top-left panel), where  we
divide our image by a 2D model from the best fitting pure elliptical isophotes.
This confirms the $c_4$ behaviour and highlights the excess of light above the
fitted elliptical model in the  central $1-7\arcsec$ of the galaxy. We associate
this with a central disk  whose major-axis coincides with that of the galaxy, at a
position angle (PA)  of $160\degr$. The presence of such a central disk was
already reported  by \citet{ss96}. There is also evidence for a secondary disk at
larger radii  ($\gtrsim12\arcsec$), revealed by an excess of light and a positive
$c_4$  parameter (Figure~\ref{Fig:centdisk}).

Another noteworthy feature is the position angle twist observed from the centre
to the outer regions of the galaxy. The magnitude of the twist is about
$5\degr$. This PA change is most likely a projection effect due to a point
symmetric structure (i.e. spiral-like structure) showing up in the outer disk, viewed at 
a large but not perfectly edge-on inclination. 

%%%%%%%%%%%%%%%%%%%%%%%%%%%%%%%%%%%%%%%%%%%%%%%%%%%%%%%%%%%%%%%%%%%%%%%%%%%%%%%%%%%%%%%%%%%%%%%%%%%%%
\section{Stellar Kinematics}
\label{Sec:Skinem}

\begin{figure}
\centering
\resizebox{\hsize}{!}{\includegraphics[angle=0]{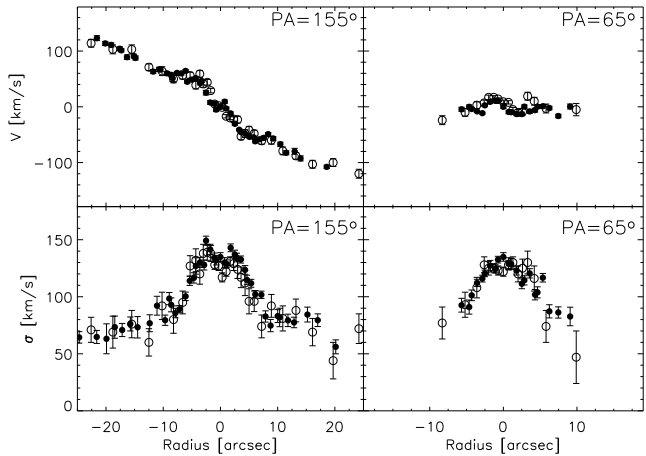}}
\caption{Comparison of NGC\,7332's kinematic profiles from \sauron\
(solid symbols) and \citeauthor{fisher94} (\citeyear{fisher94}; open
symbols) along two position angles (PA$=155\degr$ and $65\degr$).}
\label{Fig:Starscomp}
\end{figure}

We measured the \sauron\ stellar kinematics of NGC\,7332 by fitting its spectra
with a linear combination of single-age, single metallicity  population models
from \citet{vazdekis99} convolved with a Gauss-Hermite expansion. The
best-fitting parameters are determined by chi-squared minimization in pixel
space using the penalized pixel fitting (pPXF) method by \citet{capem04}, 
which is implemented as part of the {\tt XSauron} software developed at
CRAL\footnote{Centre de Recherche Astronomique de Lyon}.  We computed the
uncertainties by means of Monte Carlo simulations ($100$  realizations) and
estimate them to be $5$~\kms\ on average for the mean velocity,  $11$~\kms\ for
the velocity dispersion and $0.03$ for $h_3$ and $h_4$ for  $S/N=60$ per pixel.
This procedure traces only errors due to sampling, spectral  range and noise, but
not those introduced by a potential template mismatch.  We however minimize
template mismatch by using an optimal template (see  Section~\ref{Sec:Optemp}
below). The stellar kinematics were also measured in  the inner arcseconds using
the STIS data and \citeauthor{vdm93}'s (\citeyear{vdm93}) code, restricting the
fit to a Gaussian without higher order terms because of the poor data quality.

%----------------------------------------------------------------------------------------------------
\subsection{Measurement of emission-free stellar kinematics}
\label{Sec:Optemp}
NGC\,7332 presents the following emission lines in the rest wavelength range
covered by \sauron\ ($4790-5350$~\AA): \hbeta\ ($4861$~\AA), \oiii\  ($4959$ and
$5007$~\AA) and the \ni\ doublet ($5197.9$~\AA\ and $5200.4$~\AA).  Caution is
thus required when deriving the stellar kinematics. To obtain a clean stellar
spectrum, we followed the detailed prescriptions described below.

We first build a library of synthetic stellar spectra with different ages and
metallicities from the library of \citet{vazdekis99}, convolved  to the \sauron\
instrumental resolution (rendered uniform over the FOV). For each target spectrum,
we then perform the following steps:

\begin{enumerate} \renewcommand{\theenumi}{(\arabic{enumi})}

\item{We measure the stellar kinematics using pPXF in the wavelength 
	range $4830-5280$~\AA, avoiding the regions with nebular emission 
	(e.g., \hbeta, \oiii\ and \ni\ lines). This procedure provides the best 
	template fit to the full wavelength range.}

\item{The subtraction of the best template fit from the original dataset results 
	in a `pure emission line' spectrum that is used to extract the gas kinematics. 
	For this, we fit the emission spectrum approximating each line with a Gaussian. 
	The gaussian fit is then removed from the original data, yielding an 
	`emission line-free' spectrum from which clean line-strength indices can be derived.}

\end{enumerate}
An illustration of this procedure for NGC\,7332 is presented in  Section~\ref{Sec:Gkinem}.

In order to check the accuracy of our results, we compared them with the best
available long-slit stellar kinematics of NGC\,7332 \citep{fisher94}. In
Figure~\ref{Fig:Starscomp}, the velocity and  velocity dispersion profiles along
two perpendicular position angles,  PA$=155\degr$ and PA$=65\degr$, are shown.
They are consistent within  the errors. 

%----------------------------------------------------------------------------------------------------
\subsection{A KDC \& central disk in NGC\,7332}
\label{Sec:CRC}
\begin{figure*}
\begin{center}
\includegraphics[angle=90,width=1.\linewidth]{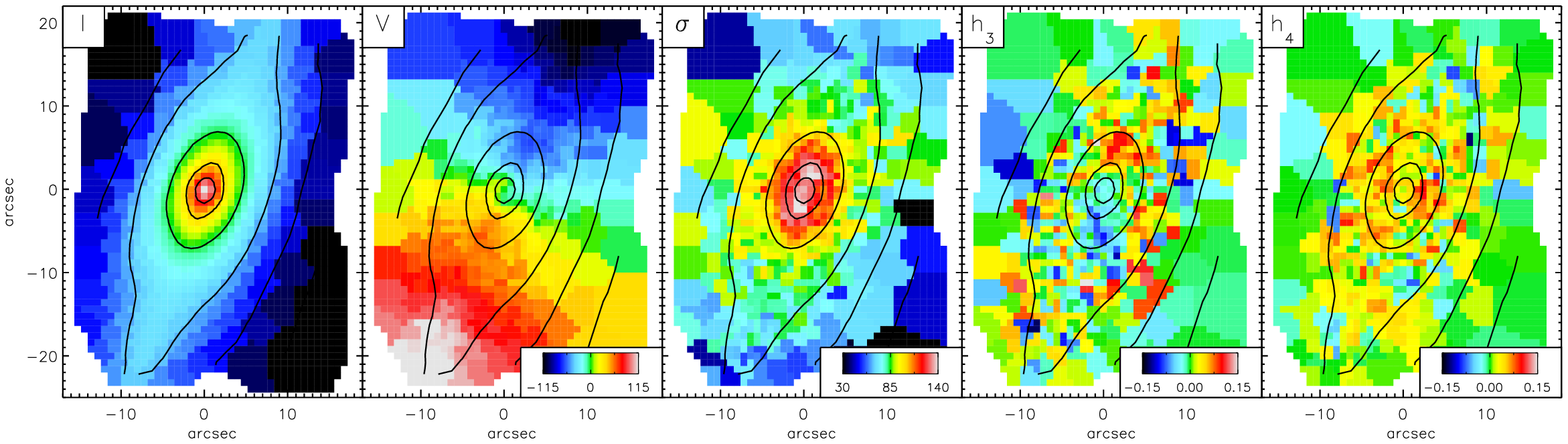}
\end{center}
\caption{Stellar kinematics of NGC\,7332 from \sauron. a) Reconstructed
  total intensity (mag~arcsec$^{-2}$, arbitrary zero point), b) mean velocity (\kms), 
  c) velocity dispersion (\kms), d) and e) Gauss-Hermite moments h$_3$ and h$_4$. 
  The \sauron\ spectra have been spatially binned to a minimum $S/N$ of
  $60$ by means of the Voronoi 2D binning algorithm of \citet{michele03}. 
  Isophotes from the reconstructed total intensity image are overlaid on 
  all maps in $1$~mag~arcsec$^{-2}$ steps.}
\label{Fig:Skin}
\end{figure*}

\begin{figure}
\centering
\includegraphics[angle=90,width=1.\linewidth]{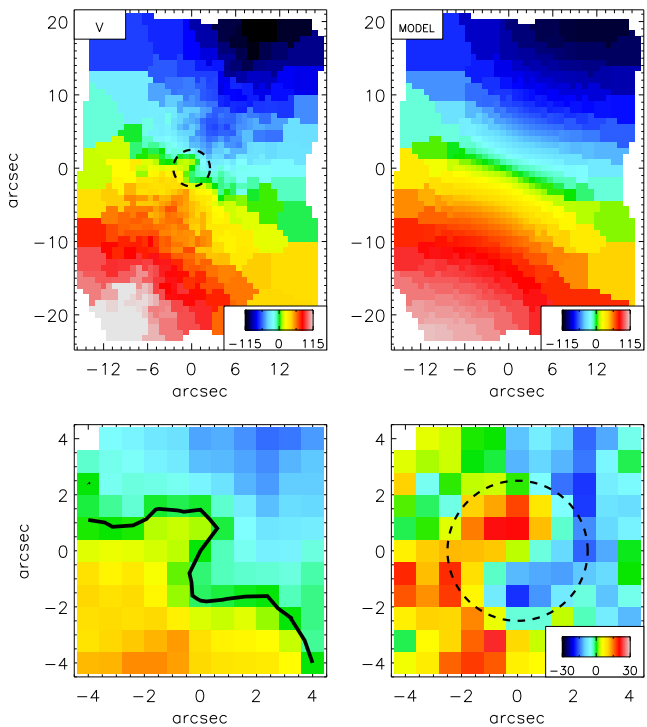}
\caption{Fit to the \sauron\ velocity field of an exponential disk model
\citep{freeman70}. Top panels: original velocity field as shown in
Figure~\ref{Fig:Skin} (left) and fitted model (right).
Bottom panels: zoom on the \sauron\ velocity field with the
zero iso-velocity curve over-imposed (solid line; left) and the same
region after subtraction of the best fit model (residual $V-V_{\rm
model}$; right). All the maps are in units of \kms. The area within the
dashed circle was excluded from the fit.}
\label{Fig:Ring}
\end{figure}

\begin{figure}
\centering
\includegraphics[angle=90,width=0.95\linewidth]{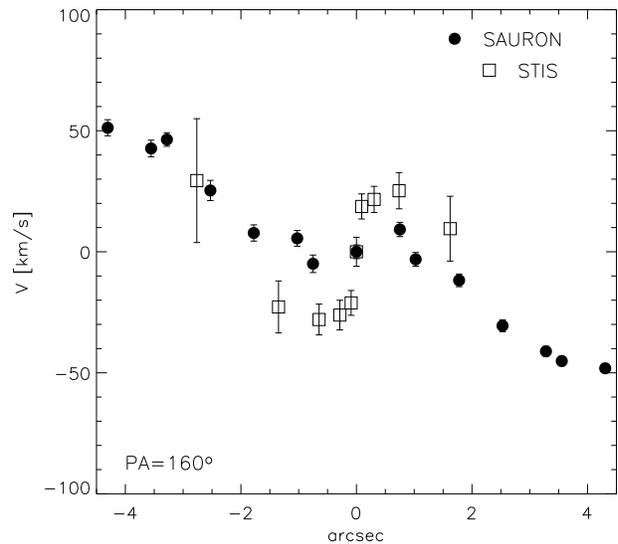}
\caption{Comparison of the \sauron\ (solid symbols) and STIS (open symbols)
velocity profiles at a position angle of $160\degr$ (the major-axis). 
The \sauron\ error bars are, in some cases, smaller than the symbols size.}
\label{Fig:STIScomp}
\end{figure}

As sown in Figure~\ref{Fig:Skin}, NGC\,7332's stellar kinematics displays a rather
smooth velocity field with rotation along the  major-axis, but it also shows some
peculiarities. One is a KDC  reported here for the first time
($\rm{r}\lesssim2\farcs5$) and revealed  by an 'S'-shaped zero-velocity contour in
the \sauron\ stellar  velocity field, emphasized in the enlarged version provided
in  Figure~\ref{Fig:Ring} (bottom left panel). We also present a comparison  of
the STIS spectroscopy and \sauron\ data along PA$\approx160\degr$  (the apparent
major-axis in the inner parts; see Fig.~\ref{Fig:STIScomp}).  At the spatial
resolution of \sauron, we only detect a weak central  counter-rotating structure,
but the signature is clear in the STIS data,  with an amplitude of about
$30$~\kms\ in the central $2\arcsec$.

In order to emphasize these central features, we have modeled the \sauron\
velocity field on large scales with a simple exponential disk \citep{freeman70},
excluding the central $2\farcs5$ from the fit to avoid being affected by the KDC.
The result is shown in Figure~\ref{Fig:Ring}. The best model was forced to have
PA$=160\degr$, consistent with the photometric major-axis as measured in
Section~\ref{Sec:Phot} and from the \sauron\ reconstructed total intensity image.
The residual image (Fig.~\ref{Fig:Ring}, bottom right panel) highlights the KDC
component: its angular momentum projected onto the minor-axis of NGC\,7332 is
negative, justifying its identification as a counter-rotating component (CRC). The
kinematic major-axis of this CRC seems however tilted by $\approx20\degr$ with
respect to that of the main galaxy.

We note that there is only very weak evidence for a flattening of the  velocity
profile in the central arcseconds along PA$=155\degr$ (see 
Fig.~\ref{Fig:Starscomp}; Section~\ref{Sec:Optemp}) and we cannot find  any
conclusive evidence for the presence of counter-rotation at this PA.  It is
however clearly seen at PA$=160$ (the major-axis; see  Fig.~\ref{Fig:STIScomp}).
And although the detection of this structure is  strongly dependent on the spatial
resolution, it again illustrates the  advantage of IFUs over long-slit
spectrographs.

\begin{figure}
\centering
\resizebox{\hsize}{!}{\includegraphics[angle=90]{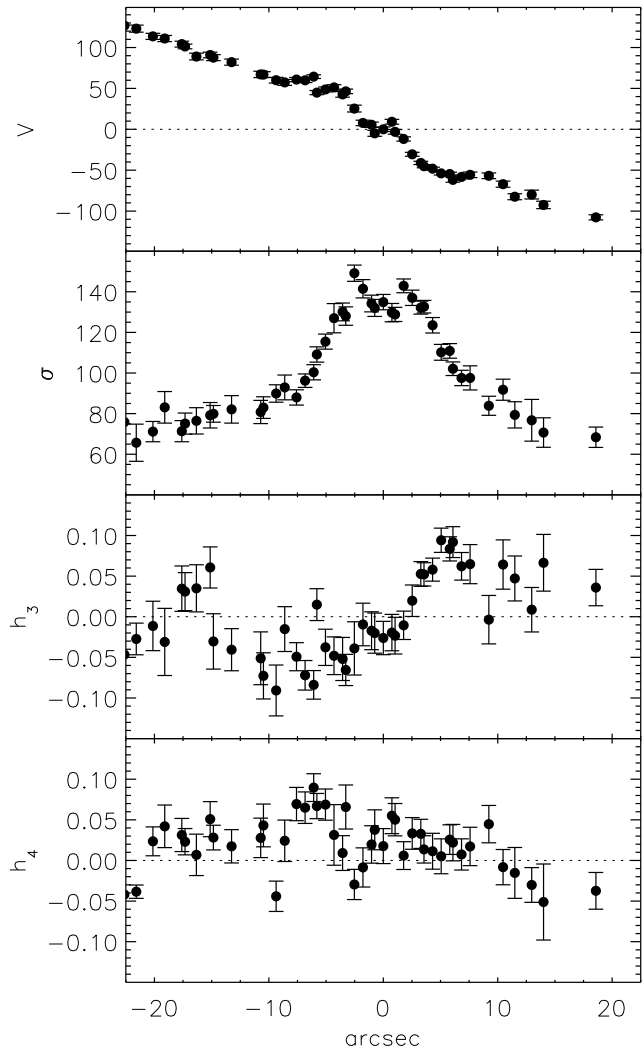}}
\caption{NGC\,7332's stellar kinematic major-axis profiles from \sauron\ 
(PA$=160\degr$). The radial velocity and velocity dispersion are in \kms.}
\label{Fig:Bar}
\end{figure}

The velocity dispersion map shown in Figure~\ref{Fig:Skin} reveals a central  dip
of about $10$~\kms\ along the major-axis of NGC\,7332 (see also 
Fig.~\ref{Fig:Bar}), in contrast with the traditional peak found at the  center of
the bulges of most early-type galaxies. This drop in velocity  dispersion is on
the same scale as the KDC ($\rm{r}\lesssim2\farcs5$; see  Fig.~\ref{Fig:Skin}) and
may well be associated with it.

In addition to the photometric evidence presented in Section~\ref{Sec:Phot}, 
there are also kinematic features supporting the presence of a central disk  on a
larger scale than the KDC. In particular, the Gauss-Hermite moment $h_3$
($\approx0.1$ in amplitude) is strongly anti-correlated with $V$ in the region
where the central disk is postulated. The superposition of a fast-rotating disk to
a moderately rotating spheroid produces a similar asymmetry in the line-of-sight
velocity distributions (LOSVDs).

We have therefore discovered a remarkably unusual situation in the core of
NGC\,7332, with a stellar KDC ($\rm{r}\lesssim2\farcs5$) and a larger central disk
($\rm{r}\lesssim7\arcsec$).

%%%%%%%%%%%%%%%%%%%%%%%%%%%%%%%%%%%%%%%%%%%%%%%%%%%%%%%%%%%%%%%%%%%%%%%%%%%%%%%%%%%%%%%%%%%%%%%%%%%%%
\section{Gas Kinematics}
\label{Sec:Gkinem}

NGC\,7332 is best known for its peculiar gas structure, namely the presence  of
two kinematically decoupled gas components, as observed by \citet{bertola92}, 
\citet{fisher94} and \citet{plana96}. \citet{fisher94} measured the \oiii\ 
emission using several slits along different position angles, whereas 
\citet{plana96} obtained complete 2D emission maps using the \halpha\ line 
(Fabry-Perot observations). Here, we are able to map three different emission 
lines within the \sauron\ wavelength range (\hbeta, \oiii, \ni).

\begin{figure*}
\centering
{\includegraphics[angle=90, width=1.\linewidth]{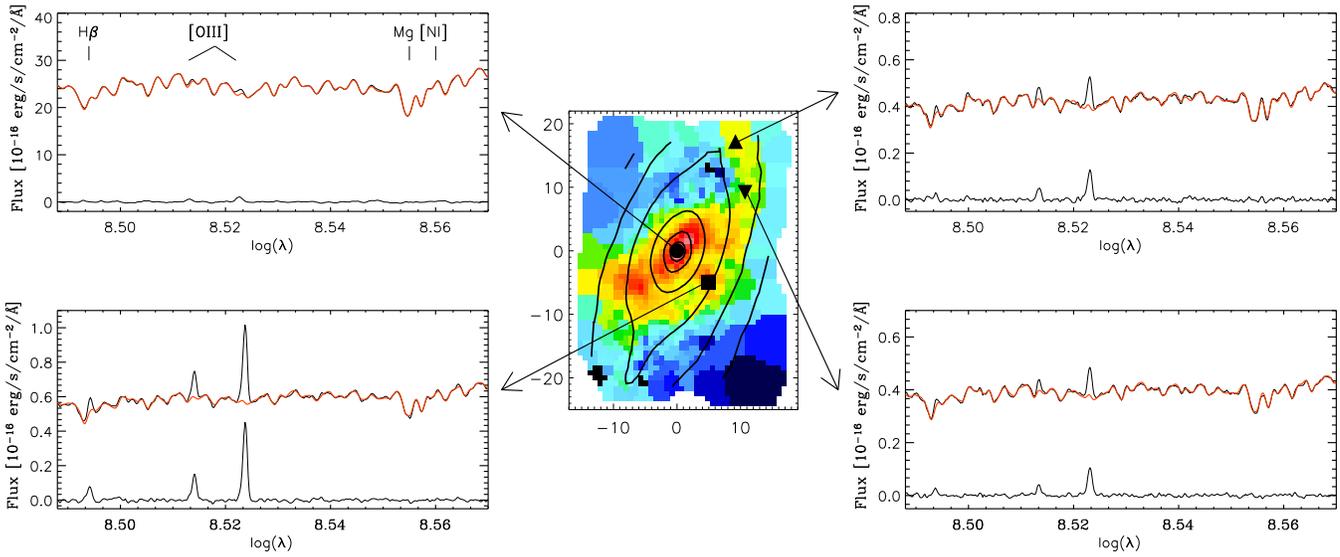}}
\caption{Optimal template fitting for NGC\,7332 in four different
  locations: the KDC (circle), main disk (upward triangle), 
  minor axis (square) and bar (downward triangle). 
  Red lines represent the best fitting stellar templates
  (see Section~\ref{Sec:Optemp}) of the underlying galaxy spectra
  (black lines). Residual spectra are shown underneath. The
  central panel shows the \oiii\ intensity distribution
  (Fig.~\ref{Fig:Gaskin}) with isophotes from the reconstructed total intensity
  image (Fig.~\ref{Fig:Skin}) overlaid in $1$~mag~arcsec$^{-2}$ steps.}
\label{Fig:Proof}
\end{figure*}

\begin{figure*}
\centering
\resizebox{\hsize}{!}{\includegraphics[angle=90]{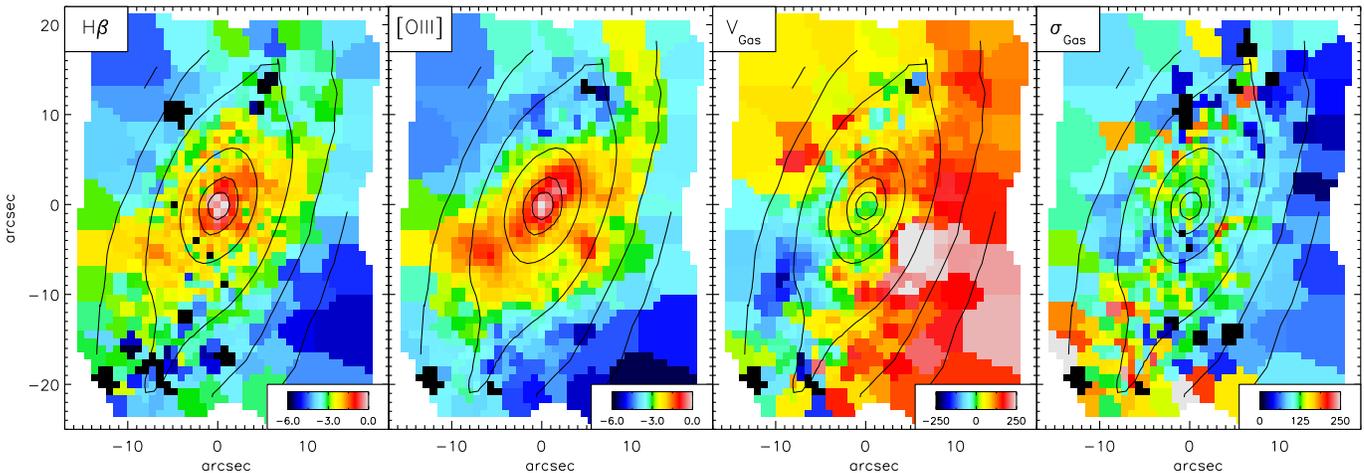}}
\caption{\hbeta\ and \oiii\ gas distribution and kinematics from
\sauron. The intensities are in mag~arcsec$^{-2}$ (arbitrary zero point) 
while the velocity and velocity dispersion are in \kms. Isophotes from
the reconstructed total intensity image (Fig.~\ref{Fig:Skin}) are
overlaid on all maps in $1$~mag~arcsec$^{-2}$ steps.}
\label{Fig:Gaskin}
\end{figure*}

We obtained a pure emission line datacube using the procedure described in
Section~\ref{Sec:Optemp}. An illustration of the method is shown in
Figure~\ref{Fig:Proof} for four different locations in the \sauron\ FOV. The
fitted synthetic spectra (red lines) are overlaid on the \sauron\ spectra of
NGC\,7332 (black lines) and the differences are shown underneath. The residuals
are small outside the spectral regions where emission lines are expected to
contribute, with oscillations just slightly above that expected from the
corresponding noise level. As shown in Figure~\ref{Fig:Gaskin}, \oiii\ is strong
everywhere in the galaxy. The \hbeta\ line, although present, is much fainter than
\oiii. We measured the \hbeta\ and \oiii\ kinematics by fitting a single Gaussian
to each of the three lines simultaneously, which relies on the assumption that all
the lines have the same velocity and velocity dispersion. We have also assumed a
1:2.96 ratio for the \oiii\  components \citep{osterbrock} but left the \oiii\ to
\hbeta\ ratio free. In our analysis, these assumptions seem to hold everywhere in
the  field. Located on the wing of the Mg\,$b$\ absorption line ($5170$~\AA),  the
\ni\ doublet is very difficult to detect. Our analysis reveals negligible  amounts
of \ni\ emission within the \sauron\ FOV so we have not corrected  the original
spectra for its presence.

In Figure~\ref{Fig:Gaskin}, we present the \hbeta\ and \oiii\ intensity maps as
well as their kinematics in the form of mean radial velocity and velocity
dispersion maps. The intensity distribution of the \hbeta\  and \oiii\ lines have
similarities, although the \hbeta\ intensity is much fainter. The detection of
\hbeta\ emission at the centre is somewhat marginal (compared to the bright
underlying stellar contribution). We find a bright counter-rotating ionised gas
component and weak traces of a   much fainter co-rotating one. We were able to
obtain \oiii\ intensity values  for the main counter-rotating component easily but
the second component  is much more difficult to isolate, not only because of its
low relative  contribution but also because of \sauron's limited spectral
resolution (see  Fig.~\ref{Fig:Proof}). We could thus only trace it in a small
region of our field, and decided not to attempt to derive its velocity field.
There is however no doubt that this second component exists, as emphasized by
\citet{fisher94} and \citet{plana96}.

The emission line maps we have obtained for NGC\,7332 show a complex structure. 
The bulk of the \oiii\ emission is clearly misaligned with respect to the main 
stellar disk and shows strong departures from an axisymmetric distribution. We 
detect a long arc-like filament in the \oiii\ distribution on the NW side of 
NGC\,7332, connecting onto the major-axis of the galaxy at the northern limit  of
the \sauron\ FOV. The velocity and velocity dispersion maps of the gaseous 
component also show strong non-axisymmetric kinematics and we find an \oiii\ 
emitting region isolated in velocity ($\approx350$~\kms\ from systemic) about 
$7\arcsec$ SW of the galaxy centre. 

Masking out all marginal \oiii\ detections (essentially all flux below $-3$ in
Fig.~\ref{Fig:Gaskin}) and the high velocity material mentioned above, the \oiii\
map then exhibits a bar or spiral-like structure within the central $10\arcsec$.
This structure is similar to, but less contrasted than, the one found in the other
nearly edge-on lenticular galaxy NGC\,3377 \citep{bacon01}. Assuming that
NGC\,7332 is  close to edge-on, at an inclination of $\approx75\degr$
\citep{apb95}, then the \oiii\ most likely has a substantial vertical extension
and is  simply seen in projection against the disk. If, on the other hand,
NGC\,7332  is more face-on, then this structure could indeed be in the equatorial
plane. 

\begin{figure*}
\centering
\includegraphics[angle=90, width=1.\linewidth]{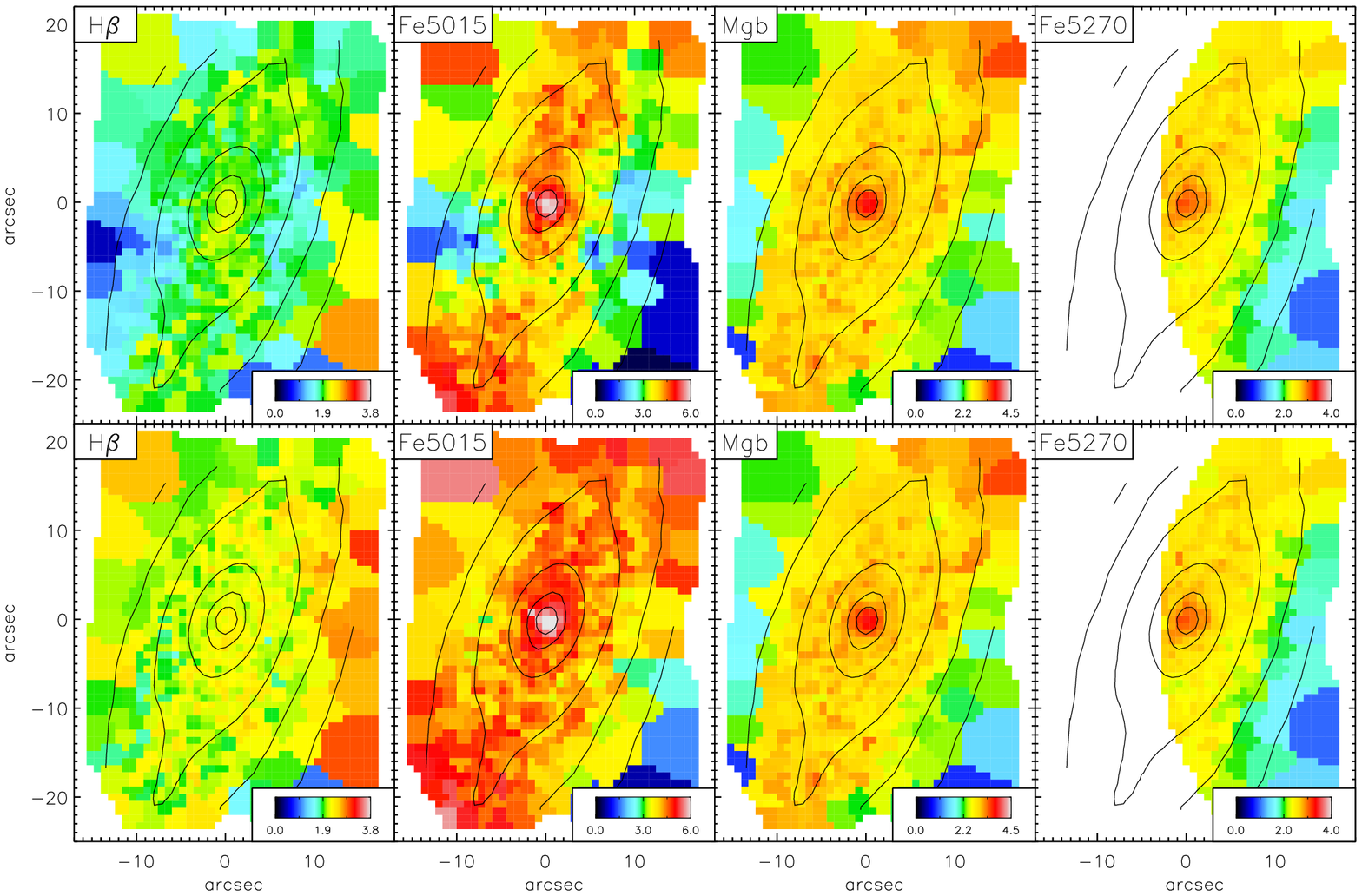}
\caption{\sauron\ \hbeta, Fe5015, Mg\,$b$\ and Fe5270 line-strength maps
on the Lick/IDS system \citep{worthey94}. Top row: raw indices. Bottom
row: emission-corrected indices. Isophotes from
the reconstructed total intensity image (Fig.~\ref{Fig:Skin}) are
overlaid on all maps in $1$~mag~arcsec$^{-2}$ steps.}
\label{Fig:Indices}
\end{figure*}

%%%%%%%%%%%%%%%%%%%%%%%%%%%%%%%%%%%%%%%%%%%%%%%%%%%%%%%%%%%%%%%%%%%%%%%%%%%%%%%%%%%%%%%%%%%%%%%%%%%%%
\section{Stellar Populations}
\label{Sec:Popu}

The many distinct kinematic components identified in the previous
sections raise the question of the sequence of events leading to the
current structure of NGC\,7332. Valuable information can be obtained by
investigating the stellar populations over the full FOV of \sauron.
The measurement of absorption line-strengths together with predictions
of stellar population models can be used to infer the
luminosity-weighted age and metallicity of the stellar populations
\citep[e.g.,][]{gonzalez93,worthey94,kd98}.

We used our flux-calibrated datacube to measure several line-strength
indices available in the \sauron\ wavelength range (i.e. \hbeta,
Fe5015, Mg\,$b$\ and Fe5270). These indices were then calibrated onto
the Lick/IDS system \citep{worthey94} following the procedures outlined
in \citet{kuntschner00}. Specifically, we observed 73 stars in common
with the Lick/IDS system to establish the small systematic offsets for
each index \citep[see also][]{wo97}. We also estimated the velocity dispersion 
corrections, to account for the broadening of the spectral features by 
the LOSVDs, using a sub-set of model spectra by \citet{vazdekis99} broadened 
to the Lick/IDS resolution.  

The Lick/IDS Fe5270 index cannot be fully mapped with \sauron\ due to
the varying bandpass over its FOV \citep{tim02}. In order to maximize the available
FOV, a new index was defined (Fe5270$_S$) which measures the same spectral 
feature, but has a reduced spectral coverage in the red pseudo-continuum band. 
This new index was then converted onto the original Lick/IDS system via the 
empirical, linear relation.

\begin{equation}
{\rm Fe5270} = 1.26 \cdot {\rm Fe5270_{S}} + 0.12
\end{equation}
The 1$\sigma$ standard deviation of this empirical calibration is $\pm$0.05\,\AA\/ 
for the Fe5270 index. More details on the \sauron\ line-strengths system 
are given in a forthcoming paper of the main \sauron\ series that applies 
to the full \sauron\ survey.

We measured the indices before and after applying our emission
subtraction procedure (Section~\ref{Sec:Optemp}). As can be seen in the
line-strength maps shown in Figure~\ref{Fig:Indices}, the corresponding
corrections are large in the outer parts due to the substantial
contamination by the emission lines (mostly \hbeta\ and \oiii). The
emission corrected line-strength values do somewhat depend on the
combination of single-age, single-metallicity population models used to
build the optimal template and fit the raw data. But although this
procedure is critical for the removal of the \hbeta\ contamination from
the \hbeta\ index, it is negligible for the removal of the \oiii\ 
contamination from the Fe5015 index, and the Mg\,$b$\ and Fe5270
indices are totally unaffected. The maximum error in the \hbeta\ index
introduced by using different sets of optimal template libraries is
$0.08$\,\AA\ ($1\sigma$).

After the emission-subtraction process, we find that \hbeta\ remains
nearly constant (within the errors) over the whole \sauron\ FOV. Since
the fitting procedure works on the spectra completely independently
from one another, it is unlikely that a flat \hbeta\ map could be
obtained by an inaccurate subtraction of the emission. The Mg\,$b$,
Fe5015 and Fe5270 maps show a steep increase in line-strength within
$5\arcsec$ of the centre (see also Fig.~\ref{Fig:LS_profiles}), but
outside of this inner region the metal line-strengths are approximately
constant over the FOV. There is, however, some evidence that the values
of the Mg\,$b$, Fe5015 and Fe5270 indices decrease slightly at $\rm
r\approx10\arcsec$ and increase again at the outermost radii along the
major axis (see Fig.~\ref{Fig:Indices} and Fig.~\ref{Fig:LS_profiles}).

The overall line-strength distribution is sufficiently uniform that
none of the galaxy components identified in the previous sections (i.e.
the KDC, inner disk, bar, and outer disks) can be clearly identified in
these maps. Specifically, the flat \hbeta\ index suggests a rather
uniform age across the entire FOV covered by \sauron. If the
(luminosity-weighted) age of the bulge or the central disk in NGC\,7332
were much different from that of the main body, it should be seen in
the \hbeta\ map (\hbeta\ is sensitive to age and independent of
metallicity changes to first order; see \citealt{worthey94})

Index-index diagrams such as \hbeta\ vs [MgFe]\footnote{ $\rmn{[MgFe]}
  \equiv \sqrt{\rmn{Mg}\,b\, \times \frac{(\rmn{Fe5270} +
      \rmn{Fe5335})}{2}}$}, together with stellar population model
predictions, are often used to estimate luminosity-weighted ages and
metallicities of galaxies. This combination of indices is almost
insensitive to non-solar abundance ratios which can otherwise
significantly affect age and metallicity estimates
\citep[e.g.,][]{tra2000,kuntschner01,TMB03}. Here, the use of the
Fe5270 and Mg\,$b$\ indices as metallicity indicators is unfortunately
limited by the restricted wavelength range of \sauron. In order to
minimize the effects of non-solar abundance ratios, we have thus
defined a new index, $\rmn{[MgFe52]}^{\prime} \equiv (0.62\times
\rmn{Mg}\,b + \rmn{Fe5270})/2.0$, using the model predictions of
\citet{TMB03}. 

In Figure~\ref{Fig:AgeMet}, we present the results of our analysis by
averaging $1\farcs6\times1\farcs6$ regions (i.e.\ $2\times2$ lenslets) 
around the four key positions shown in Figure~\ref{Fig:Proof}. The
estimated luminosity-weighted age for all regions is $5\pm2$~Gyr. The
metallicity of  the central area is about $2$ times solar while the
outer areas show  metallicities between $0.35$ and $1$ times solar.
These results are  consistent with others found in the literature. Using
a similar line-strength  analysis, the nucleus of NGC\,7332 has been
reported to host a population of  luminosity-weighted age between $4$
and $6$~Gyr depending on the authors  \citep{vazari99,terlevich02}. The
color gradients are also small in the disk  and bulge \citep{pb97},
consistent with our finding of an approximately  constant age across the
\sauron\ FOV.

The young luminosity-weighted age of $5\pm2$~Gyr detected in our
analysis clearly shows that NGC\,7332 has experienced relatively recent
star formation. Furthermore, the uniformity of the \hbeta\ map suggests 
that this star formation affected all regions of the galaxy probed by
our integral-field instrument. 

%%%%%%%%%%%%%%%%%%%%%%%%%%%%%%%%%%%%%%%%%%%%%%%%%%%%%%%%%%%%%%%%%%%%%%%%%%%%%%%%%%%%%%%%%%%%%%%%%%%%%
\section{Discussion}
\label{Sec:Discussion}

In this paper, we studied a galaxy that has sometimes been labelled peculiar.
Using integral-field spectroscopy, we were able to better constrain the stellar
kinematics, the ionised gas distribution and kinematics, and the stellar
populations of NGC\,7332. In turn, those can now be used to better constrain its
evolutionary history. A summary of some key numbers extracted from the data is 
given in Table~\ref{Tab:SAURON_Params}.

\subsection{The presence of a bar}
\label{Sec:Bar}

A number of photometric features described in Section~\ref{Sec:Phot} are
consistent with NGC\,7332 being a barred galaxy viewed close to edge-on.
Indeed, soon after they form, bars buckle and settle with an increased thickness,
appearing boxy-shaped when seen end-on (i.e.\  along the bar major-axis) and
peanut-shaped when seen side-on (i.e.\ along the bar minor-axis; see e.g.\
\citealt{cdfp90} and \citealt{rsjk91} for N-body simulations;
\citealt{merrifield99} and \citealt{bureau99b} for observations). The boxiness of
the isophotes does not show properly in the $c_4$ profile of
Figure~\ref{Fig:centdisk} because the isophotes are simultaneously boxy (at large
galactic heights) and disky (along the equatorial plane), due to the superposition
of the boxy bar and an inner disk. The boxiness is however obvious in the image
shown in Figure~\ref{Fig:FoV}. The plateau in the major-axis surface brightness
profile at intermediate radii found by \citet[][]{ss96} and confirmed by our
$I$-band image (Fig.~\ref{Fig:FoV}, right panel) is also characteristic  of bars
viewed nearly edge-on \citep[see e.g.,][]{ldp00,ba04}. NGC\,7332  finally exhibits
a double disk structure, which can be interpreted as the  result of secular
evolution due to the presence of a bar (see e.g. \citealt{vdbemsellem98} and
references therein).

\begin{figure}
\centering
\includegraphics[angle=90, width=1.\linewidth]{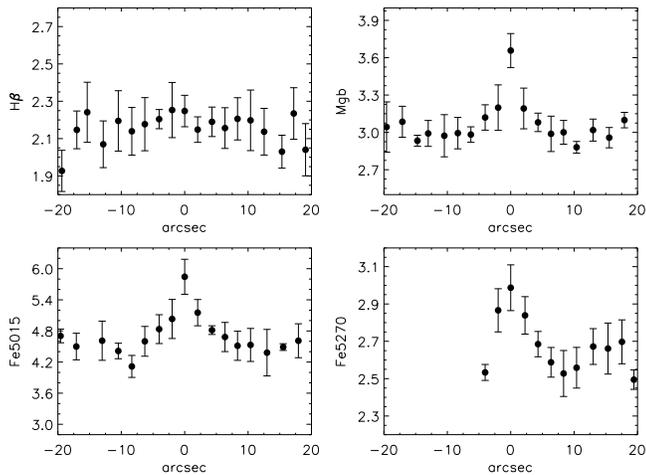}
\caption{\sauron\ emission-corrected line-strength major-axis profiles 
(\hbeta, Mg\,$b$, Fe5015 and Fe5270) on the Lick/IDS system \citep{worthey94}. 
The error bars represent Poisson errors.}
\label{Fig:LS_profiles}
\end{figure}

Although NGC\,7332 is close to edge-on, it is still possible to use its
(projected) kinematics to verify the existence of the large-scale thick bar
suggested by the photometry (see Section~\ref{Sec:Phot}; \citealt{ba04}).  The
extracted major-axis kinematic profiles of NGC\,7332 (Fig.~\ref{Fig:Bar};  but see
also \citealt{fisher94}) show kinematic features commonly seen in  early-type
barred galaxies: a 'double-hump' velocity profile and a central  (local) minimum
in the velocity dispersion. Double-hump velocity profiles are observed in
dissipationless simulations of barred disks \citep{ba04} and in a number of
galaxies with  so-called 'double-disk' structures \citep[e.g.,][]{ss96,bba98},
also often associated with barred systems \citep[e.g.,][]{vdbemsellem98}. Central
velocity dispersion minima have also been observed in barred galaxies
\citep[e.g.,][]{emsellem01b} and  interpreted as the signature of a cold stellar
disk, presumably resulting from  recent (bar-driven) gas accumulation at the
center \citep{emsellem01b,wozniak03}.  The very weak dependence of rotation on
galactic height shown in  Figure~\ref{Fig:Skin} is also consistent with NGC\,7332
being barred \citep[e.g.,][]{cdfp90,am02}, although axisymmetric configurations
can also in principle give rise to cylindrical rotation \citep[e.g.,][]{r88}.

The only major difference with the kinematic bar signatures emphasized by 
\citet{ba04} from N-body simulations is the strong $h_3-V$ anti-correlation  in
the central parts ($\rm{r}\lesssim7\arcsec$). This is however commonly observed  in
boxy bulges otherwise showing strong evidence for the presence of a bar 
\citep{cb04} and most likely results from the absence of a dissipative  component
in the simulations \citep[see][]{fb93,wozniak03}.

The rather homogeneous stellar population of NGC\,7332 is also consistent with the
presence of bar. Indeed, the so-called bulge and disk are then  both made-up of
the same (disk) material, and any population gradient that  may have been present
before the bar formed or may develop afterwards will  be smoothed out by the large
radial and vertical motions of the stars, at  least within the bar region
\citep[e.g.,][]{martin94,friedli94}. Gradients in the very centre may however
survive or develop due to bar-driven inflows.

The presence of a bar in NGC\,7332 was already hinted at by \citet{fisher94},  who
were however left to speculate about its existence. \citet{ldp00} estimated  a
diameter of $56\arcsec$ for the (weak) bar in NGC\,7332 via the characteristic 
major-axis plateau (observed here in the near-infrared). Considering the 
discussion above, we can now assert its presence with more confidence. For the 
remainder of this discussion, we will thus take for granted the presence of a  bar
in NGC\,7332 and focus our efforts on understanding its origin and role in  the
evolution of the galaxy.

\begin{table}
\begin{center}
\caption{\sauron\ parameters for NGC\,7332.}
\label{Tab:SAURON_Params}
\begin{tabular}{lll}
\hline
Parameter                       & Value & Units\\
\hline
Stellar Heliocentric Systemic Velocity	&	1206 $\pm$ 5		& \kms   \\
Gas Heliocentric Systemic Velocity		& 1206$^a$ $\pm$ 7		& \kms   \\
Stellar V$_{\rm max}$					& 123$^b$ $\pm$ 5		& \kms   \\
Extent of the KDC						& $\sim$2.5				& arcsec \\
Extent of the Central Disk				& $\sim$7.0				& arcsec \\
Mass of counter-rotating ionised gas	& $\sim$1.5				& $10^5$ \Msun \\
EW(\hbeta)								& 2.23$^c$ $\pm$ 0.09	& \AA  \\
EW(Mg\,$b$)								& 3.45$^c$ $\pm$ 0.18	& \AA  \\
EW($\rmn{[MgFe52]}^{\prime}$)			& 2.53$^c$ $\pm$ 0.07	& \AA  \\
\hline
\end{tabular}
\end{center}
$^a$ from the \hbeta\ and \oiii\ lines fitted simultaneously.\\
$^b$ within the \sauron\/ field.\\
$^c$ central aperture [1\farcs6 x 1\farcs6]
\end{table}

\subsection{Origin and status of the gas}
\label{Sec:Origin_gas}

NGC\,7332 exhibits some amount of counter-rotating ionised gas ($\approx 1.5 \cdot
10^{5}$ \Msun), which has commonly been used  to argue for an external origin. In
the case of NGC\,7332, the ionised gas most likely comes from a tidal interaction
with the neighbouring  galaxy NGC\,7339. We note that extended counter-rotating
ionised gas is  common in S0s \citep{bertola92}, and more specifically in galaxies
with  boxy bulges (Bureau \& Chung 2004, in preparation). 

\begin{figure}
\centering
{\includegraphics[angle=90,width=0.99\linewidth]{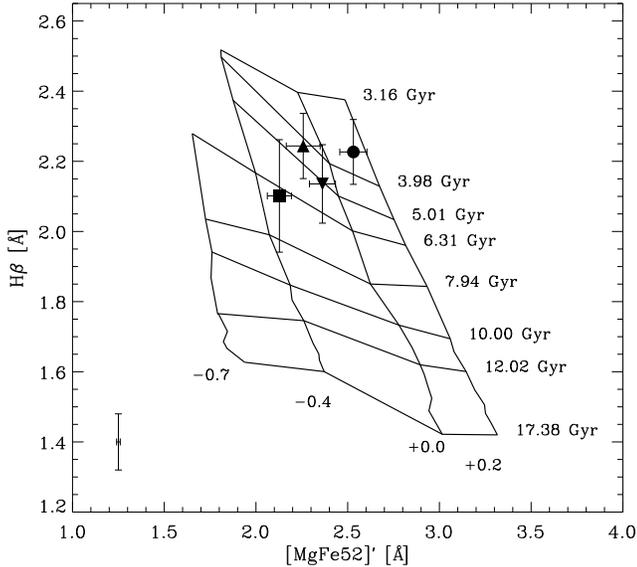}}
\caption{\hbeta\ vs $\rmn{[MgFe52]}^{\prime}$ equivalent width diagram. 
  Symbols represent measurements from the clean datacube at the four
  positions shown in Figure~\ref{Fig:Proof}: the KDC (circle), main disk 
  (upward triangle), minor axis (square) and bar (downward triangle).
  The error bar in the bottom left corner shows the $1\sigma$ uncertainty 
  introduced by the emission subtraction process. The error bars shown 
  for the data points represent Poisson errors.}
\label{Fig:AgeMet}
\end{figure}

Although the ionised gas distribution and kinematics obtained from the \sauron\
datacube are complex, some of the counter-rotating gaseous  component in NGC\,7332
seems to have a relatively ordered flow and extends  relatively far from the
major-axis of the galaxy. If we assume that the  galaxy is sufficiently inclined,
the ionised gas could trace a bar or  spiral-like structure in the equatorial
plane. If quasi edge-on, however,  it must have a significant vertical extent.
Considering the orbital timescales  involved (e.g.\ about $2.5\times10^6$~yr for a
gas clump $1$~kpc from the  centre), we would naively expect the (dissipative)
counter-rotating ionised  gas to be quickly driven back to the equatorial plane.
This need not be so, however, as in the tumbling potential of a bar there exists
one major family of stable retrograde orbits (the 'anomalous' orbits) which is 
tilted with respect to the equatorial plane and can reach significant  heights
\citep{pfenniger91,friedli93,emsellem97}. 

Despite the above ambiguity, it is safe to assume that an accretion event  is at
the origin of the observed counter-rotating gaseous component in  NGC\,7332. The
same accretion event could be responsible for the formation  of the central KDC
revealed by \sauron\ and STIS, as well as the dip in  the velocity dispersion
profile (see Section~\ref{Sec:CRC}) and the increase  of the Mg\,$b$ and Fe5270
indices at the very centre (see Section~\ref{Sec:Popu}).  The \hbeta\ index values
would then setup an upper limit of about 3-4 Gyrs  (luminosity-weighted) for the
KDC (see Fig.~\ref{Fig:AgeMet}). Another later  accretion event may be required to
explain the more perturbed gas clumps  located further away from the equatorial
plane (see Section~\ref{Sec:Gkinem}),  which have rather high relative velocities.
This gas has certainly not  settled yet, but it will eventually fall back towards
the equatorial plane of  NGC\,7332.

%---------------------------------------------------------------------------------
\subsection{Evolutionary scenarios}
\label{Sec:Evolution}

Any formation and evolution scenario for NGC\,7332 must take into account the two
main ingredients hinted at from the observations  described in this paper: the
presence of a bar and a recent interaction  event. We now examine them in turn.

\subsubsection{The role of the bar and the central stellar disk}

In our view, the main morphological and kinematic features that we associate with
the bar in Section~\ref{Sec:Bar} do not represent  a proper illustration of
bar-driven evolution processes. This is  because, although those properties become
more extreme as the bar  strengthens, most of them are established on timescales
equivalent to the bar formation itself  \citep[e.g.,][]{cdfp90,rsjk91,ba04}. In
the following paragraphs, we  thus concentrate on (presumably) bar-driven
processes which take place  gradually over a much longer timescale (i.e.\ tens of
dynamical times).

In the context of the bar hypothesis, the interaction between the natural
frequencies of the system (e.g., the angular and epicyclic frequencies of the
circular orbits) and the pattern speed of the bar implies the existence of
resonances, around which the galaxy is shaping. The photometric and kinematic
features we observe in NGC\,7332 will thus likely be associated with those
resonances.  The double disk structure is particularly interesting in this respect
since it likely builds up gradually through bar-driven processes 
\citep{vdbemsellem98}.

The inner disk is usually interpreted as marking the region within the Inner
Lindblad Resonance (ILR). The major-axis plateau observed by \citet{ldp00} extends
to a radius of about $25\arcsec$ ($\approx2.8$~kpc),  setting a lower limit for
the length of the bar. Assuming the standard  corotation radius to bar length
ratio $r_{\rm CR}/r_{\rm b}=1.2$ for  early-type barred galaxies (e.g.,
\citealt{a92}; \citealt{gkm03} and  references therein), $r_{\rm
CR}\approx30\arcsec$. A simple model then  predicts the ILR to be at roughly
$12\arcsec$ (see \citealt{vdbemsellem98}),  indeed marking the transition between
the inner disk and the outer disk.  This may be supported by the line-strength
index maps, where there is evidence  for a weak decrease of the Fe5015, Mg$b$ and
Fe5270 profiles at about  $10\arcsec$ along the major-axis (see
Figs.~\ref{Fig:Indices} and  \ref{Fig:LS_profiles}).

A comparison with the edge-on S0 galaxy NGC\,4570 appears appropriate here,  since
it may be at a different stage of a similar (bar-driven) process.  NGC\,4570 was
studied in detail by \citet[][]{vdbemsellem98}, who identified  similar
morphological features (i.e.\ double disk and intermediate boxy  structures) to
those in NGC\,7332. Their study also revealed the same trends  in ellipticity and
boxiness ($c_4$). The case for a bar was  made there almost exclusively based on
the identification of two edge-on  rings, exceptionally well characterized and
consistent with the locations of  bar resonances. The structures observed in
NGC\,4570 are however one order of magnitude smaller than in NGC\,7332: the
'nuclear' disk is less than  $1\arcsec$ in radius and the inner ring is at a
radius of $\approx1\farcs7$. The  main difference between the two galaxies then
lies in the detection of a  large-scale bar of about $3$~kpc in NGC\,7332, whereas
\citet{vdbemsellem98}  hinted at the presence of an 'inner' bar of $\approx320$~pc
in NGC\,4570 (with no  evidence of a large-scale bar). Another obvious qualitative
difference is the  presence of a significant amount of counter-rotating stars and
gas in the  central region of NGC\,7332. We could speculate on the existence of a
secondary  bar in NGC\,7332, but its detection would require high spatial
resolution  photometry with high signal-to-noise (e.g.\ HST WFPC-2 or ACS images).
It may  thus well be that NGC\,7332 is simply in an earlier stage of a gradual
bar-driven evolutionary process.

\subsubsection{The role of interactions}

The formation of decoupled cores is often mentioned within the hierarchical
structure formation framework \citep[e.g.,][]{kgw94}, involving multiple mergers
to form galaxies. As mentioned above, although there is no evidence for a recent
major merger, the gas distribution and kinematics strongly suggest that
NGC\,7332 was recently involved in an interaction event (most likely with
NGC\,7339). The KDC could then correspond to the last or one of the last major
accretion (and star formation) events  in such a hierarchy
\citep[e.g.,][]{davies01}. In fact, the bar itself could be the result of that 
same event, as bar are easily excited by tidal interactions 
\citep[e.g.,][]{noguchi87,gerin90}.

The neighbour galaxy NGC\,7339, a Sb galaxy placed 5\arcmin\ away from NGC\,7332, 
is most likely the one responsible for such an event. The position of this 
galaxy is such that the offplane gas clump found in NGC\,7332 could well be a direct continuation 
of the west side of NGC\,7339's disk. Additionally, the maximum receding velocity 
measured on the west side of this galaxy is $\sim$1525 \kms \citep{courteau97}, which 
is strikingly similar to the radial velocity of the high velocity clump in NGC\,7332 
($\sim$1550 \kms). This picture leads to the suggestion that this clump is linked 
to NGC\,7339. Deep H{\sc i} observations could confirm this idea.

It is hard to assess which of the bar or the interaction occurred first, and 
whether the KDC was formed from material funneled by the former or the latter. 
NGC\,7332 will, however, continue to evolve on a timescale of a few $10^8$~yr. 
As emphasized above, the gas will settle down due to its dissipative nature, and 
some of it may feed the inner few arcseconds (with or without more star formation). 
In this context, NGC\,7332 may be seen as a precursor of the now gas-poor edge-on 
S0 galaxy NGC\,4570. 

%%%%%%%%%%%%%%%%%%%%%%%%%%%%%%%%%%%%%%%%%%%%%%%%%%%%%%%%%%%%%%%%%%%%%%%%%%%%%%%%%%%%%%%%%%%%%%%%%%%%%
\section{Conclusions}
\label{Sec:Conclusions}

We presented in this paper new \sauron\ integral-field spectroscopy observations
of the edge-on S0 galaxy NGC\,7332, along with a discussion of existing ground and
space-based imaging.  

The photometric analysis reveals a boxy bulge, a central disk and  evidence for
the presence of a bar (Section~\ref{Sec:Phot}; but see  also \citealt{ldp00}). The
\sauron\ observations provide unprecedented  coverage of the stellar and gas
kinematics. In particular, the stellar  kinematics displays a rather smooth
velocity field with rotation along  the major-axis and a weak dependence of
rotation on galactic height. We  also discovered a stellar kinematically decoupled
component (KDC) misaligned  with respect to the galaxy's kinematic major-axis,
which may be related  to a dip in the centre of the velocity dispersion map
($\rm{r}\lesssim2\farcs5$). We  finally provided kinematic evidence for the
presence of a larger central  stellar disk ($\rm{r}\lesssim7\arcsec$). 

NGC\,7332 presents significant ionised gas emission in the spectral range 
studied, in particular in the \hbeta\ ($4861$~\AA) and \oiii\ ($4959$ and 
$5007$~\AA) lines. We used a sophisticated technique to separate the  absorption
and emission lines, yielding both pure absorption spectra from  which to derive
the stellar kinematics and pure emission spectra for the gas  kinematics (and
distribution). The emission maps reveal a very complex gas  morphology and
kinematics. This is found especially in \oiii\, which is  mainly counter-rotating
with respect to the stars. The gas maps  (Figure~\ref{Fig:Gaskin}) show
significant amounts of gas at high relative  velocities outside of the equatorial
plane, whereas the central parts display  rather ordered motions. We also found
traces of a faint co-rotating ionised gas  component.

The analysis of the absorption line-strengths in NGC\,7332 show that its stellar
populations are generally young ($5\pm2$~Gyr), not only in the disk but also in
the bulge, in agreement with previous studies \citep{bp94,vazdekis96,terlevich02}.
The metallicity indices (i.e.\ Fe5015,  Mg$b$, Fe5270) show an increase in the
centre, contrasting with the  rather homogeneous \hbeta\ index. There is also weak
evidence of a  decrease at $r\approx10$\arcsec, between the central and the outer 
disks.

The existence of a large-scale bar in NGC\,7332 can simultaneously explain  most
of the features found in the galaxy. The boxy morphology can be  explained by the
thickness of the bar after buckling, which also leads  to homogeneous stellar
populations across the bulge and disk. The stellar  kinematics can similarly be
explained by the particular orbital structure  of barred disks
\citep[see][]{ba04}. The KDC, on the other hand, was most  likely formed through
late gas infall, an accretion event perhaps related  to the complex ionised gas
distribution currently seen.

As emphasized by the unique data set presented in this paper, the S0 galaxy 
NGC\,7332 possesses a number of peculiar properties helping to unravel  its
formation and evolution processes. There is strong observational evidence  that
interactions played a role in the shaping of NGC\,7332, while the bar  has and is
certainly significantly affecting its evolution. Given the amount  of gas yet to
settle and the relatively young age of the stellar populations  in NGC\,7332, it
probably represents an earlier evolutionary stage of more  regular edge-on S0
galaxies (i.e.\ NGC\,4570, \citealt{vdbemsellem98};  NGC\,3115, \citealt{eric03};
NGC\,3377, \citealt{bacon01}). The morphology  and dynamics of present-day S0
galaxies can thus probably be understood  through the combined effects of both
interactions and bar-driven processes.

%%%%%%%%%%%%%%%%%%%%%%%%%%%%%%%%%%%%%%%%%%%%%%%%%%%%%%%%%%%%%%%%%%%%%%%%%%%%%%%%%%%%%%%%%%%%%%%%%%%%%
\section*{Acknowledgements}
We would like to thank Fabien Wernli for stimulating discussions in the first
stages of this work. The William Herschel Telescope is  operated on the island of
La Palma by the Isaac Newton Group in the Observatorio del Roque de los
Muchachos of the Instituto de Astrof\'\i sica  de Canarias. Support for this work
was provided by NASA through Hubble Fellowship grant HST-HF-01136.01 awarded by
the Space Telescope Science Institute, which is operated by the Association of
Universities for Research in Astronomy, Inc., for NASA, under contract
NAS~5-26555. MC acknowledges support from a VENI grant awarded by the Netherlands
Organization of Scientific Research (NWO).

%%%%%%%%%%%%%%%%%%%%%%%%%%%%%%%%%%%%%%%%%%%%%%%%%%%%%%%%%%%%%%%%%%%%%%%%%%%%%%%%%%%%%%%%%%%%%%%%%%%%%


\begin{thebibliography}{}

\bibitem[\protect\citeauthoryear{{Andredakis}, {Peletier} \&
  {Balcells}}{{Andredakis} et~al.}{1995}]{apb95}
{Andredakis} Y.~C.,  {Peletier} R.~F.,    {Balcells} M.,  1995, \mnras, 275,
  874

\bibitem[\protect\citeauthoryear{{Athanassoula}}{{Athanassoula}}{1992}]{a92}
{Athanassoula} E.,  1992, \mnras, 259, 345

\bibitem[\protect\citeauthoryear{{Athanassoula} \& {Misiriotis}}{{Athanassoula}
  \& {Misiriotis}}{2002}]{am02}
{Athanassoula} E.,  {Misiriotis} A.,  2002, \mnras, 330, 35

\bibitem[\protect\citeauthoryear{{Bacon et~al.}}{{Bacon et~al.}}{2001}]{bacon01}
{Bacon} R. et~al.,  2001, \mnras, 326, 23

\bibitem[\protect\citeauthoryear{{Baggett}, {Baggett} \& {Anderson}}
 {{Baggett} et~al.}{1998}]{bba98}
{Baggett} W.~E.,  {Baggett} S.~M., {Anderson} K.S.J.,  1998, \aj, 116, 1626

\bibitem[\protect\citeauthoryear{{Balcells} \& {Peletier}}{{Balcells} \&
  {Peletier}}{1994}]{bp94}
{Balcells} M.,  {Peletier} R.~F.,  1994, \aj, 107, 135

\bibitem[\protect\citeauthoryear{{Bender}}{{Bender}}{1988}]{bender88}
{Bender} R.,  1988, \aap, 202, L5

\bibitem[\protect\citeauthoryear{{Bertola}, {Buson} \& {Zeilinger}}{{Bertola}
  et~al.}{1992}]{bertola92}
{Bertola} F.,  {Buson} L.~M.,    {Zeilinger} W.~W.,  1992, \apjl, 401, L79

\bibitem[\protect\citeauthoryear{{Bureau} \& {Athanassoula}}{{Bureau} \&
  {Athanassoula}}{2004}]{ba04}
{Bureau} M.,  {Athanassoula} E.,  2004, submitted to \apj

\bibitem[\protect\citeauthoryear{{Bureau} \& {Freeman}}{{Bureau} \&
  {Freeman}}{1999}]{bureau99b}
{Bureau} M.,  {Freeman} K.~C.,  1999, \aj, 118, 126

\bibitem[\protect\citeauthoryear{{Caldwell}}{{Caldwell}}{1983}]{caldwell83}
{Caldwell} N.,  1983, \apj, 268, 90

\bibitem[\protect\citeauthoryear{{Cappellari} \& {Copin}}{{Cappellari} \&
  {Copin}}{2003}]{michele03}
{Cappellari} M.,  {Copin} Y.,  2003, \mnras, 342, 345

\bibitem[\protect\citeauthoryear{{Cappellari} \& {Emsellem}}{{Cappellari} \&
  {Emsellem}}{2004}]{capem04}
{Cappellari} M.,  {Emsellem} E.,  2004, \pasp, 116, 138

\bibitem[\protect\citeauthoryear{{Cappellari}, {Verolme}, {van der Marel},
  {Verdoes}, {Illingworth}, {Franx}, {Carollo} \& {de Zeeuw}}{{Cappellari}
  et~al.}{2002}]{michele02}
{Cappellari} M.,  {Verolme} E.~K.,  {van der Marel} R.~P.,  {Verdoes} G.~A.~V.,
  {Illingworth} G.~D.,  {Franx} M.,  {Carollo} C.~M.,    {de Zeeuw} P.~T.,
  2002, \apj, 578, 787

\bibitem[\protect\citeauthoryear{{Cardiel}}{{Cardiel}}{1999}]{cardiel99}
{Cardiel} N.,  1999, Ph.D.~Thesis, Univ. Complutense de Madrid

\bibitem[\protect\citeauthoryear{{Carollo}}{{Carollo}}{1999}]{carollo99}
{Carollo} C.~M.,  1999, \apj, 523, 566

\bibitem[\protect\citeauthoryear{{Carter}}{{Carter}}{1978}]{carter78}
{Carter} D.,  1978, \mnras, 182, 797

\bibitem[\protect\citeauthoryear{{Chung} \& {Bureau}}{{Chung} \&
  {Bureau}}{2004}]{cb04}
{Chung} A.,  {Bureau} M.,  2004, submitted to \aj

\bibitem[\protect\citeauthoryear{{Combes}, {Debbasch}, {Friedli} \&
  {Pfenniger}}{{Combes} et~al.}{1990}]{cdfp90}
{Combes} F.,  {Debbasch} F.,  {Friedli} D.,    {Pfenniger} D.,  1990, \aap,
  233, 82

\bibitem[\protect\citeauthoryear{{Courteau}}{{Courteau}}{1997}]{courteau97}
{Courteau} S., 1997, \aj, 114, 2404

\bibitem[\protect\citeauthoryear{{Courteau}, {de Jong} \& {Broeils}}{{Courteau}
  et~al.}{1996}]{courteau96}
{Courteau} S.,  {de Jong} R.~S.,    {Broeils} A.~H.,  1996, \apjl, 457, L73

\bibitem[\protect\citeauthoryear{{Davies et~al.}}{{Davies et~al.}}{2001}]{davies01}
{Davies} R.~L. et~al.,  2001, \apjl, 548, L33

\bibitem[\protect\citeauthoryear{{de Vaucouleurs}}{{de
  Vaucouleurs}}{1959}]{devac59}
{de Vaucouleurs} G.,  1959, Handbuch der Physik, 53, 275

\bibitem[\protect\citeauthoryear{{de Vaucouleurs}, {de Vaucouleurs}, {Corwin},
  {Buta}, {Paturel} \& {Fouque}}{{de Vaucouleurs} et~al.}{1991}]{RC3}
{de Vaucouleurs} G.,  {de Vaucouleurs} A.,  {Corwin} H.~G.,  {Buta} R.~J.,
  {Paturel} G.,    {Fouque} P.,  1991, \skytel, 82, 621

\bibitem[\protect\citeauthoryear{{de Zeeuw et~al.}}{{de Zeeuw et~al.}}{2002}]{tim02}
{de Zeeuw} P.~T. et~al.,  2002, \mnras, 329, 513

\bibitem[\protect\citeauthoryear{{Dressler}}{{Dressler}}{1980}]{dressler80}
{Dressler} A.,  1980, \apj, 236, 351

\bibitem[\protect\citeauthoryear{{Emsellem}}{{Emsellem}}{2003}]{eric03}
{Emsellem} E.,  2003, in Carnegie Observatories Astrophysics Series, Vol. 1:
  Coevolution of Black Hole and Galaxies, ed L.C. Ho Cambridge University Press

\bibitem[\protect\citeauthoryear{{Emsellem} \& {Arsenault}}{{Emsellem} \&
  {Arsenault}}{1997}]{emsellem97}
{Emsellem} E.,  {Arsenault} R.,  1997, \aap, 318, L39

\bibitem[\protect\citeauthoryear{{Emsellem}, {Greusard}, {Combes}, {Friedli},
  {Leon}, {P{\' e}contal} \& {Wozniak}}{{Emsellem} et~al.}{2001}]{emsellem01b}
{Emsellem} E.,  {Greusard} D.,  {Combes} F.,  {Friedli} D.,  {Leon} S.,  {P{\'
  e}contal} E.,    {Wozniak} H.,  2001, \aap, 368, 52

\bibitem[\protect\citeauthoryear{{Emsellem et al.}}{{Emsellem et al.}}{2004}]{emsellem04}
{Emsellem}, E. et~al., 2004, submitted to \mnras

\bibitem[\protect\citeauthoryear{{Faber} \& {Gallagher}}{{Faber} \&
  {Gallagher}}{1976}]{faber76b}
{Faber} S.~M.,  {Gallagher} J.~S.,  1976, \apj, 204, 365

\bibitem[\protect\citeauthoryear{{Fisher}}{{Fisher}}{1997}]{fisher97}
{Fisher} D.,  1997, \aj, 113, 950

\bibitem[\protect\citeauthoryear{{Fisher}, {Illingworth} \& {Franx}}{{Fisher}
  et~al.}{1994}]{fisher94}
{Fisher} D.,  {Illingworth} G.,    {Franx} M.,  1994, \aj, 107, 160

\bibitem[\protect\citeauthoryear{{Franx} \& {Illingworth}}{{Franx} \&
  {Illingworth}}{1988}]{fi88}
{Franx} M.,  {Illingworth} G.~D.,  1988, \apjl, 327, L55

\bibitem[\protect\citeauthoryear{{Freeman}}{{Freeman}}{1970}]{freeman70}
{Freeman} K.~C.,  1970, \apj, 160, 811

\bibitem[\protect\citeauthoryear{{Friedli} \& {Benz}}{{Friedli} \& {Benz}}{1993}]{fb93}
{Friedli} D.,  {Benz} W.,  1993, \aap, 268, 65

\bibitem[\protect\citeauthoryear{{Friedli} \& {Udry}}{{Friedli} \&
  {Udry}}{1993}]{friedli93}
{Friedli} D.,  {Udry} S.,  1993, in DeJonghe, H., Habing,~H.~J., eds, 
Proc. IAU Symp, 153: Galactic Bulges, Kluwer Academic Publishers, Dordrecht, p.~273

\bibitem[\protect\citeauthoryear{{Friedli}, {Benz} \& {Kennicutt}}{{Friedli}, 
{Benz} \& {Kennicutt}}{1994}]{friedli94}
{Friedli} D.,  {Benz} W.,  {Kennicutt} R., 1994, \apjl, 430, L105

\bibitem[\protect\citeauthoryear{{Gerin}, {Combes} \&
{Athanassoula}}{{Gerin}, {Combes} \& {Athanassoula}}{1990}]{gerin90}
{Gerin} M.,  {Combes} F.,  {Athanassoula} E.,  1990, \aap, 320, 37

\bibitem[\protect\citeauthoryear{{Gerssen}, {Kuijken} \&
{Merrifield}}{{Gerssen}, {Kuijken} \& {Merrifield}}{2003}]{gkm03}
Gerssen J.,  Kuijken K.,  Merrifield M. R.,  2003, \mnras, 345, 261

\bibitem[\protect\citeauthoryear{{Gonz{\' a}lez}}{{Gonz{\'
  a}lez}}{1993}]{gonzalez93}
{Gonz{\' a}lez} J.~J., 1993, Ph.D.~Thesis, Univ. of California Santa Cruz

\bibitem[\protect\citeauthoryear{{Goudfrooij}, {Baum} \& {Walsh}}{{Goudfrooij}
  et~al.}{1997}]{goud97}
{Goudfrooij} P.,  {Baum} S.~A.,    {Walsh} J.~R.,  1997, The 1997 HST
  Calibration Workshop with a New Generation of Instruments, p.~100

\bibitem[\protect\citeauthoryear{{Gunn} \& {Gott}}{{Gunn} \&
  {Gott}}{1972}]{gunn72}
{Gunn} J.~E.,  {Gott} J.~R.~I.,  1972, \apj, 176, 1

\bibitem[\protect\citeauthoryear{{Hubble}}{{Hubble}}{1936}]{hubble36}
{Hubble} E.~P.,  1936, The Realm of the Nebulae, Yale University Press

\bibitem[\protect\citeauthoryear{{Kauffmann}, {Guiderdoni} \&
  {White}}{{Kauffmann} et~al.}{1994}]{kgw94}
{Kauffmann} G.,  {Guiderdoni} B.,    {White} S.~D.~M.,  1994, \mnras, 267, 981

\bibitem[\protect\citeauthoryear{{Kuntschner}}{{Kuntschner}}{2000}]{kuntschner00}
  {Kuntschner} H., 2000, \mnras, 315, 184

\bibitem[\protect\citeauthoryear{{Kuntschner} \& {Davies}}{{Kuntschner} \&
  {Davies}}{1998}]{kd98}
{Kuntschner} H.,  {Davies} R.~L.,  1998, \mnras, 295, L29

\bibitem[\protect\citeauthoryear{{Kuntschner}, {Lucey}, {Smith}, {Hudson} \&
  {Davies}}{{Kuntschner} et~al.}{2001}]{kuntschner01}
{Kuntschner} H.,  {Lucey} J.~R.,  {Smith} R.~J.,  {Hudson} M.~J.,    {Davies}
  R.~L.,  2001, \mnras, 323, 615

\bibitem[\protect\citeauthoryear{{L{\" u}tticke}, {Dettmar} \& {Pohlen}}{{L{\"
  u}tticke} et~al.}{2000}]{ldp00}
{L{\" u}tticke} R.,  {Dettmar} R.-J.,    {Pohlen} M.,  2000, \aap, 362, 435

\bibitem[\protect\citeauthoryear{{Larson}, {Tinsley} \& {Caldwell}}{{Larson}
  et~al.}{1980}]{larson80}
{Larson} R.~B.,  {Tinsley} B.~M.,    {Caldwell} C.~N.,  1980, \apj, 237, 692

\bibitem[\protect\citeauthoryear{{Lauer}}{{Lauer}}{1995}]{lauer95}
{Lauer} T.~R. e.~a.,  1995, \aj, 110, 2622

\bibitem[\protect\citeauthoryear{{Martin} \& {Roy}}{{Martin} \&
{Roy}}{1994}]{martin94}
Martin P., Roy J.-R., 1994, \apj, 424, 599

\bibitem[\protect\citeauthoryear{{Merrifield} \& {Kuijken}}{{Merrifield} \&
  {Kuijken}}{1999}]{merrifield99}
{Merrifield} M.~R.,  {Kuijken} K.,  1999, \aap, 345, L47

\bibitem[\protect\citeauthoryear{{Noguchi}}{{Noguchi}}{1987}]{noguchi87}
{Noguchi}, M.,  1987, \mnras, 228, 635 


\bibitem[\protect\citeauthoryear{{Osterbrock}}{{Osterbrock}}{1989}]{osterbrock}
{Osterbrock} D.~E.,  1989, Astrophysics of Gaseuous Nebulae and Active Galactic
Nuclei, Mill Valley, CA, University Science Books, 1989, p.~422 

\bibitem[\protect\citeauthoryear{{Peletier} \& {Balcells}}{{Peletier} \&
  {Balcells}}{1996}]{pb96}
{Peletier} R.~F.,  {Balcells} M.,  1996, \aj, 111, 2238

\bibitem[\protect\citeauthoryear{{Peletier} \& {Balcells}}{{Peletier} \&
  {Balcells}}{1997}]{pb97}
{Peletier} R.~F.,  {Balcells} M.,  1997, New Astronomy, 1, 349

\bibitem[\protect\citeauthoryear{{Peletier}, {Balcells}, {Davies},
  {Andredakis}, {Vazdekis}, {Burkert} \& {Prada}}{{Peletier}
  et~al.}{1999}]{peletier99}
{Peletier} R.~F.,  {Balcells} M.,  {Davies} R.~L.,  {Andredakis} Y.,
  {Vazdekis} A.,  {Burkert} A.,    {Prada} F.,  1999, \mnras, 310, 703

\bibitem[\protect\citeauthoryear{{Pfenniger}}{{Pfenniger}}{1993}]{pfenniger93}
{Pfenniger} D.,  1993, DeJonghe, H., Habing,~H.~J., eds, Proc. IAU Symp, 153: 
Galactic Bulges, Kluwer Academic Publishers, Dordrecht, p.~387

\bibitem[\protect\citeauthoryear{{Pfenniger} \& {Friedli}}{{Pfenniger} \&
  {Friedli}}{1991}]{pfenniger91}
{Pfenniger} D.,  {Friedli} D.,  1991, \aap, 252, 75

\bibitem[\protect\citeauthoryear{{Plana} \& {Boulesteix}}{{Plana} \&
  {Boulesteix}}{1996}]{plana96}
{Plana} H.,  {Boulesteix} J.,  1996, \aap, 307, 391

\bibitem[\protect\citeauthoryear{{Raha}, {Sellwood}, {James} \& {Kahn}}{{Raha}
  et~al.}{1991}]{rsjk91}
{Raha} N.,  {Sellwood} J.~A.,  {James} R.~A.,    {Kahn} F.~D.,  1991, \nat,
  352, 411

\bibitem[\protect\citeauthoryear{{Rowley}}{{Rowley}}{1988}]{r88}
{Rowley} G.,  1988, \apj, 331, 124

\bibitem[\protect\citeauthoryear{{Sandage}}{{Sandage}}{1961}]{sandage61}
{Sandage} A.,  1961, Washington: The Hubble Atlas of Galaxies, Carnegie Institution.

\bibitem[\protect\citeauthoryear{{Sandage} \& {Visvanathan}}{{Sandage} \&
  {Visvanathan}}{1978}]{sandage78}
{Sandage} A.,  {Visvanathan} N.,  1978, \apj, 225, 742

\bibitem[\protect\citeauthoryear{{Schweizer}}{{Schweizer}}{1986}]{schweizer86}
{Schweizer} F.,  1986, Science, 231, 227

\bibitem[\protect\citeauthoryear{{Seifert} \& {Scorza}}{{Seifert} \&
  {Scorza}}{1996}]{ss96}
{Seifert} W.,  {Scorza} C.,  1996, \aap, 310, 75

\bibitem[\protect\citeauthoryear{{Sellwood} \& {Wilkinson}}{{Sellwood} \&
  {Wilkinson}}{1993}]{sw93}
{Sellwood} J.~A.,  {Wilkinson} A.,  1993, Reports on Progress in Physics, 56,
  173

\bibitem[\protect\citeauthoryear{{Statler} \& {Smecker-Hane}}{{Statler} \&
  {Smecker-Hane}}{1999}]{ss99}
{Statler} T.~S.,  {Smecker-Hane} T.,  1999, \aj, 117, 839

\bibitem[\protect\citeauthoryear{{Terlevich} \& {Forbes}}{{Terlevich} \&
  {Forbes}}{2002}]{terlevich02}
{Terlevich} A.~I.,  {Forbes} D.~A.,  2002, \mnras, 330, 547

\bibitem[\protect\citeauthoryear{{Terndrup}, {Davies}, {Frogel}, {Depoy} \&
  {Wells}}{{Terndrup} et~al.}{1994}]{terndrup94}
{Terndrup} D.~M.,  {Davies} R.~L.,  {Frogel} J.~A.,  {Depoy} D.~L.,    {Wells}
  L.~A.,  1994, \apj, 432, 518

\bibitem[\protect\citeauthoryear{{Thomas}, {Maraston} \& {Bender}}{{Thomas}
  et~al.}{2003}]{TMB03}
{Thomas} D.,  {Maraston} C.,    {Bender} R.,  2003, \mnras, 339, 897

\bibitem[\protect\citeauthoryear{{Tonry}, {Dressler}, {Blakeslee}, {Ajhar},
  {Fletcher}, {Luppino}, {Metzger} \& {Moore}}{{Tonry} et~al.}{2001}]{tonry01}
{Tonry} J.~L.,  {Dressler} A.,  {Blakeslee} J.~P.,  {Ajhar} E.~A.,  {Fletcher}
  A.,  {Luppino} G.~A.,  {Metzger} M.~R.,    {Moore} C.~B.,  2001, \apj, 546, 681

\bibitem[\protect\citeauthoryear{{Toomre}}{{Toomre}}{1977}]{toomre77}
{Toomre} A.,  1977, \araa, 15, 437

\bibitem[\protect\citeauthoryear{{Trager}, {Faber}, {Worthey} \& {Gonz{\'
  a}lez}}{{Trager} et~al.}{2000}]{tra2000}
{Trager} S.~C.,  {Faber} S.~M.,  {Worthey} G.,    {Gonz{\' a}lez} J.~J.,  2000,
  \aj, 119, 1645

\bibitem[\protect\citeauthoryear{{van den Bergh}}{{van den
  Bergh}}{1960a}]{vdbergh60a}
{van den Bergh} S.,  1960a, \apj, 131, 558

\bibitem[\protect\citeauthoryear{{van den Bergh}}{{van den
  Bergh}}{1960b}]{vdbergh60b}
{van den Bergh} S.,  1960b, \apj, 131, 215

\bibitem[\protect\citeauthoryear{{van den Bosch} \& {Emsellem}}{{van den Bosch}
  \& {Emsellem}}{1998}]{vdbemsellem98}
{van den Bosch} F.~C.,  {Emsellem} E.,  1998, \mnras, 298, 267

\bibitem[\protect\citeauthoryear{{van der Marel} \& {Franx}}{{van der Marel} \&
  {Franx}}{1993}]{vdm93}
{van der Marel} R.~P.,  {Franx} M.,  1993, \apj, 407, 525

\bibitem[\protect\citeauthoryear{{Vazdekis}}{{Vazdekis}}{1999}]{vazdekis99}
{Vazdekis} A.,  1999, \apj, 513, 224

\bibitem[\protect\citeauthoryear{{Vazdekis} \& {Arimoto}}{{Vazdekis} \&
  {Arimoto}}{1999}]{vazari99}
{Vazdekis} A.,  {Arimoto} N.,  1999, \apj, 525, 144

\bibitem[\protect\citeauthoryear{{Vazdekis}, {Casuso}, {Peletier} \&
  {Beckman}}{{Vazdekis} et~al.}{1996}]{vazdekis96}
{Vazdekis} A.,  {Casuso} E.,  {Peletier} R.~F.,    {Beckman} J.~E.,  1996,
  \apjs, 106, 307

\bibitem[\protect\citeauthoryear{{Wagner}, {Bender} \& {Moellenhoff}}{{Wagner}
  et~al.}{1988}]{wbm88}
{Wagner} S.~J.,  {Bender} R.,    {Moellenhoff} C.,  1988, \aap, 195, L5

\bibitem[\protect\citeauthoryear{{Wernli}, {Emsellem} \& {Copin}}{{Wernli}
  et~al.}{2002}]{wernli02}
{Wernli} F.,  {Emsellem} E.,    {Copin} Y.,  2002, \aap, 396, 73

\bibitem[\protect\citeauthoryear{{White} \& {Rees}}{{White} \&
  {Rees}}{1978}]{wr78}
{White} S.~D.~M.,  {Rees} M.~J.,  1978, \mnras, 183, 341

\bibitem[\protect\citeauthoryear{{Worthey}, {Faber}, {Gonzalez} \&
  {Burstein}}{{Worthey} et~al.}{1994}]{worthey94}
{Worthey} G.,  {Faber} S.~M.,  {Gonzalez} J.~J.,    {Burstein} D.,  1994,
  \apjs, 94, 687
  
\bibitem[\protect\citeauthoryear{{Worthey} \& {Ottaviani}}{{Worthey} \&
    {Ottaviani}}{1997}]{wo97}{Worthey}, G., {Ottaviani}, D.~L., 1997, \apjs, 111, 377

\bibitem[\protect\citeauthoryear{{Wozniak}, {Combes}, {Emsellem} \&
  {Friedli}}{{Wozniak} et~al.}{2003}]{wozniak03}
{Wozniak} H.,  {Combes} F.,  {Emsellem} E.,    {Friedli} D.,  2003, \aap, 409, 469

\end{thebibliography}
\end{document}